\documentclass[useAMS,usenatbib,usegraphicx]{mn2e}

\newcommand{\hii}{H\,{\sc ii}}
\newcommand{\hi}{H\,{\sc i}}
\newcommand{\ha}{H{$\alpha$}}

\newcommand{\nodata}{...}
\newcommand{\lum}{erg~s$^{-1}$}

\newcommand{\cden}{cm$^{-2}$}
\newcommand{\vden}{cm$^{-3}$}

\newcommand{\msun}{M$_{\sun}$}
\newcommand{\kms}{km~s$^{-1}$}
\newcommand\phn{\phantom{x}}
\newcommand\phs{\phantom{-}}

\newcommand\aj{AJ}
\newcommand\apj{ApJ}%
\newcommand\apjl{ApJ}%
\newcommand\apjs{ApJS}%
\newcommand\aap{A\&A}%
\newcommand\aaps{A\&AS}%
\newcommand\mnras{MNRAS}%
\newcommand\nat{Nature}%
\title[{\it Chandra} survey of dwarf starburst galaxies II]
{A {\it Chandra} X-ray survey of nearby dwarf starburst galaxies:
II. Starburst properties and outflows}

\author[J{\"u}rgen Ott, Fabian Walter, and Elias Brinks]{J{\"u}rgen Ott$^{1}$\thanks{Bolton Fellow, E-mail:
Juergen.Ott@csiro.au}, Fabian Walter$^{2}$\thanks{E-mail: walter@mpia.de},  and Elias Brinks\thanks{E-mail: ebrinks@star.herts.ac.uk}$^{3,4}$\\
$^{1}$CSIRO Australia Telescope National Facility, Cnr Vimiera \& Pembroke Roads, Marsfield NSW 2112, Australia\\
$^{2}$Max--Planck--Institut f{\"u}r Astronomie, K{\"o}nigstuhl 17, 69117 Heidelberg, Germany\\
$^{3}$Instituto Nacional de Astrof{\'{\i}}sica {\'O}ptica y Electr{\'o}nica, Apartado Postal 51 y 216, Puebla, Pue., 72000, Mexico\\
$^{4}$Centre for Astrophysics Research, University of Hertfordshire, College Lane, Hatfield AL10 9AB, England }

\begin{document}


\pagerange{\pageref{firstpage}--\pageref{lastpage}} \pubyear{2005}

\maketitle

\label{firstpage}

\begin{abstract}

We present a comprehensive comparison of the X-ray properties of a
sample of eight dwarf starburst galaxies observed with {\it Chandra}
(I\,Zw\,18, VII\,Zw\,403, NGC\,1569, NGC\,3077, NGC\,4214, NGC\,4449,
NGC\,5253, He\,2--10). In Paper\,I we presented in detail the data
reduction and analysis of the individual galaxies. For the unresolved
X-ray sources we find the following: point sources are in general
located close to bright \hii\ regions, rims of superbubbles, or young
stellar clusters. The number of X-ray point sources appears to be a
function of the current star formation rate and the blue luminosity of
the hosts. Ultraluminous X-ray sources are only found in those dwarf
galaxies which are currently interacting. The power law index of the
combined cumulative X-ray point source luminosity function is
$\alpha=0.24\pm0.06$, shallower than that of more massive starburst
galaxies ($\alpha=0.4-0.8$) and of non-starburst galaxies
($\alpha\sim1.2$). For those galaxies showing extended X-ray emission
(6 out of the 8 galaxies), we derive the following: Superwinds develop
along the steepest gradient of the \hi\ distribution with volume
densities of $0.02-0.06$\,\vden, pressures of $1-3\times
10^{5}$\,K\,cm$^{-3}$, thermal energies of $2-30\times10^{54}$\,erg,
and hot gas masses of $2-20\times10^{6}$\,\msun\ ($\sim 1$ per cent of
the \hi\ masses.
On global scales, the distribution of the X-ray emission looks
remarkably similar to that seen in H$\alpha$ (comparing azimuthal
averages); locally however their distribution is clearly distinct in
many cases -- this can be explained by the different emission
mechanisms (forward vs.\ reverse shocks). Mass-loading of order 1 to 5
is required to explain the differences between the amount of hot gas
and and the modelled mass-loss from massive stars. The metallicity of
the dwarf galaxies correlates with the diffuse X-ray luminosity and
anti-correlates with the cooling time of the hot gas. The diffuse
X-ray luminosity is also a function of the current star formation
rate. The mechanical luminosities of the developing superwinds are
energetic enough to overcome the gravitational potentials of their
host galaxies. This scenario is supported by the overpressures of the
hot gas compared to the ambient ISM. Extended \hi\ envelopes such as
tidal tails, however, may delay outflows on timescales exceeding those
of the cooling time of the hot gas.

\end{abstract}


\section{Introduction}
\label{sec:intro}

Starburst phases in galaxies dump a huge amount of mechanical and
thermal energy into the interstellar medium (ISM) of their hosts via
stellar winds and supernovae \citep[SNe; see, e.g., the models
of][]{lei99}. The mechanical energy which is deposited creates shocks
in the ambient medium: a forward shock ionises the ISM to temperatures
of $\sim 10^{4-5}$\,K and is visible as diffuse emission in \ha\ and
other optical spectral lines. A reverse shock thermalises the stellar
ejecta to temperatures of $\sim 10^{6-8}$\,K which gives rise to
X-rays \citep[e.g., ][]{cas75,wea77,hec02,str02}. This coronal gas is
characterised by high pressures which are capable of pushing material
to the halo. If the gravitational potential is shallow and the cooler
gas in the halo is not very extended, the material is able to
ultimately overcome the host's gravitational potential and the
hydrodynamical drag of the surrounding gas. Eventually, the hot gas
escapes into the intergalactic medium
\citep[][]{mac99,fer00,sil01}. To first order, the end products of
massive star formation (SF) are oblivious to their environment and
release the same amount of energy whether SF has taken place in a
spiral or in a dwarf galaxy. Dwarf galaxies, however, have a much
shallower gravitational potential and outflows can be established more
easily. As metals are produced in high mass stars, the hot gas carries
freshly produced heavy elements with it and distributes them over the
entire galaxy and/or the IGM
\citep[e.g., ][]{sil01b}. In particular, the energy input is mainly
provided by type II SNe, which predominantly release $\alpha$
elements. Iron is efficiently produced in type Ia SNe but they occur
with a substantial time delay \citep[see, e.g,][]{mat01}. For this
reason, the $\alpha/Fe$ ratio in the outflow is predicted to be higher
than solar.

This picture has three important implications. First of all,
it could explain why dwarf galaxies have a much lower metallicity
\citep[$Z\simeq 1/10$ solar, see, e.g., ][]{ski89} as compared to
larger spirals ($Z\simeq$ solar). Secondly, if a starburst is strong
enough, it can blow away \emph{all} of its ISM. In that case, the
galaxy is not able to further form stars and eventually
fades. Thirdly, the gas which is expelled to the intergalactic medium
(IGM) pollutes it with heavy elements. This might have been especially
important at large look-back times. In cold dark matter cosmologies
spiral galaxies form by mergers of smaller constituents. Dwarf
galaxies formed relatively early and may have quickly polluted the IGM
by galactic superwinds before larger systems got assembled. Due to the
larger number of small systems and a higher merging rate which
triggers SF, the removal of metal-enriched gas in dwarf galaxies by
the energetic input of SNe and stellar winds may have been more common
in the past and may have influenced the formation and evolution of
galaxies in general.

This prompted us to embark on a detailed study with the {\it Chandra
X-ray Observatory} (CXO) of a sample of eight dwarf galaxies that are
undergoing a starburst. The aim of our study is to try to trace the
hot gas in these systems, determine its extent, measure the relative
abundance of $\alpha$ elements over iron, and assess the likelihood
that this hot gas will escape the galaxy. In \citet{ott05} (hereafter
referred to as Paper\,I) we presented the sample a detailed
description of our data reduction techniques. We detected 55
unresolved sources which are likely related to the galaxies and we
presented fits to their individual spectra. Six of the dwarf galaxies
have diffuse X-ray emission and we derived such fundamental parameters
as their X-ray luminosity and the temperature of the gas.

This paper is dedicated to a comparison of the X-ray characteristics
of the dwarf galaxies in our sample and of any
correlations of their X-ray with other properties. In
Sect.\,\ref{sec:sample_props_other} we summarise the general
properties of the galaxies in our sample. This is followed  in
Sect.\,\ref{sec:morph} by a discussion of the 
morphologies of the different components (X-ray, optical, \ha, and
\hi). In Sect.\,\ref{sec:point_pop} the
point source population in the galaxies is discussed according to
their locations in the galaxies, their spectral properties and their
correlations with SF tracers. In this section we also present our 
interpretation of the point source luminosity function. The X-ray
properties of the diffuse, coronal gas are evaluated and compared in
Sect.\,\ref{sec:xrayprop}. Correlations of those properties with
quantities derived at other wavelengths are presented in
Sect.\,\ref{sec:correl} followed by a discussion on the probability of
outflows in Sect.\,\ref{sec:out}. Sect.\,\ref{sec:summary} summarises
our results.

\section{General properties of the sample galaxies}
\label{sec:sample_props_other}

\begin{table*}
\centering
\begin{minipage}{270cm}

\caption{Distances and optical properties of the galaxy sample.}
\begin{tabular}{@{}llllllll@{}}
\hline

\multicolumn{1}{c}{Galaxy}&\multicolumn{1}{c}{$D$}&\multicolumn{1}{c}{$m_{\rm B}^{0}$}&\multicolumn{1}{c}{$m_{\rm K_{s}}$}&\multicolumn{1}{c}{$d_{\rm 25}$}&\multicolumn{1}{c}{$F_{\rm H\alpha}$}&\multicolumn{1}{c}{$12+\log(O/H)$}&\multicolumn{1}{c}{References}\\
\multicolumn{1}{c}{}&\multicolumn{1}{c}{[Mpc]}&\multicolumn{1}{c}{[mag]}&\multicolumn{1}{c}{[mag]}&\multicolumn{1}{c}{[arcmin]}&\multicolumn{1}{c}{[erg~s$^{-1}$~cm$^{-2}$]}&\multicolumn{1}{c}{}&\\
\multicolumn{1}{c}{(1)}&\multicolumn{1}{c}{(2)}&\multicolumn{1}{c}{(3)}&\multicolumn{1}{c}{(4)}&\multicolumn{1}{c}{(5)}&\multicolumn{1}{c}{(6)}&\multicolumn{1}{c}{(7)}&\multicolumn{1}{c}{(8)}\\

\hline

I\,Zw\,18       &      12.6   &$    16.47\pm  0.17   $&\multicolumn{1}{c}{$\nodata$}      &$  0.41 \pm  0.11 $&$3.26\times 10^{-13}  $  &$  7.16          $ &1, 2, -, 2, 3, 4       \\
VII\,Zw4\,403    &  \phn4.5    &$    15.74\pm  0.38   $&\multicolumn{1}{c}{$\nodata$}      &$  0.93 \pm  0.13 $&$1.95\times 10^{-11}  $  &$  7.73\pm  0.01 $ &5, 2, -, 2, 6, 7      \\
NGC\,1569       &  \phn2.2    &$ \phn8.29\pm  0.32   $&$7.86\pm0.02$  &$  3.24 \pm  0.11 $&$8.28\times 10^{-11}  $ &$  8.22\pm  0.07 $ &8, 2, 9, 2, 10, 11    \\
NGC\,3077       &  \phn3.6    &$    10.19\pm  0.11   $&$7.30\pm0.02$  &$  4.90 \pm  0.60 $&$3.75\times 10^{-12}  $ &$  8.90          $ &12, 2, 9, 2, 13, 11    \\

NGC\,4214       &  \phn2.9    &$ 10.17\pm  0.21   $   &$7.91\pm0.05$  &$  8.91 \pm  1.32 $&$1.47\times 10^{-11}  $ &$  8.28\pm  0.08 $ & 14, 2, 9, 2, 15, 16   \\

NGC\,4449       &  \phn3.9    &$\phn 9.65\pm  0.62   $&$7.25\pm0.04$  &$  5.62 \pm  0.29 $&$2.03\times 10^{-11}  $ &$  8.31\pm  0.07 $ &17, 2, 9, 2, 10, 11    \\
NGC\,5253       &  \phn3.3    &$\phn9.77\pm   0.40   $&$8.29\pm0.03$  &$  4.79 \pm  0.11 $&$ 1.70\times 10^{-11} $ &$  8.23\pm  0.01 $ &18, 2, 9, 2, 19, 11    \\
He\,2-10       &  \phn9.0    &$    11.97\pm  0.01   $&$9.00\pm0.02$  & $  1.82 \pm  0.12 $&$3.50\times 10^{-12}  $ &$  8.93          $ &20, 2, 9, 2, 21, 22    \\
\hline

\end{tabular}

\label{tab:sample_props_opt}
\end{minipage}

\flushleft {\sc References:} (1) \citet{ost00}; (2) \citet{pat97}; (3) \citet{can02}; (4) \citet{gus00}; (5) \citet{lyn98}; (6) \citet{sil02}; (7) \citet{itz97}; (8) \citet{isr88}; (9) 2MASS Atlas \citet{jar03}; (10) \citet{hun93}; (11) \citet{mar97}; (12) \citet{fre94}; (13) \citet{wal02}; (14) \citet{mai02}; (15) \citet{mar98}; (16) \citet{kob96}; (17) \citet{hun98}; (18) \citet{gib00}; (19) \citet{marl97}; (20) \citet{vac92}; (21) \citet{joh00}; (22) \citet{kob99}

\end{table*}

\begin{table*}
\begin{minipage}{170mm}
\centering
\caption{\hi\ and far-infrared properties of the galaxy
sample.}
\begin{tabular}{@{}llllll@{}}
\hline

\multicolumn{1}{c}{Galaxy} & \multicolumn{1}{c}{$F_{\rm HI}$} & \multicolumn{1}{c}{$S_{\rm 60\mu}$} & \multicolumn{1}{c}{$S_{\rm 100\mu}$} & \multicolumn{1}{c}{$v_{\rm hel}$} & \multicolumn{1}{c}{References}\\
\multicolumn{1}{c}{} & \multicolumn{1}{c}{[Jy~km~s$^{-1}$]} & \multicolumn{1}{c}{[Jy]} & \multicolumn{1}{c}{[Jy]} & \multicolumn{1}{c}{[\kms\ ]} & \\
\multicolumn{1}{c}{(1)}&\multicolumn{1}{c}{(2)}&\multicolumn{1}{c}{(3)}&\multicolumn{1}{c}{(4)}&\multicolumn{1}{c}{(5)}&\multicolumn{1}{c}{(6)}\\

\hline

I\,Zw\,18      & $ \phn\phn1.19 $          &\multicolumn{1}{c}{\nodata}&\multicolumn{1}{c}{\nodata}&$\phs752\pm  8 $      &1,$-$,$-$, 2 \\ 
VII\,Zw\,403   &$ \phn14.41 $              &$\phn0.38\pm0.04 $         &$\phn0.90\pm 0.17 $    &$\phn-93\pm  6 $      &3, 4, 4, 2  \\
NGC\,1569      &$  \phn96.30\pm17.50 $     & $46.48\pm5.11 $           &$ 51.71 \pm 10.86 $        &$\phn-90\pm 11 $             &5, 4, 4, 2 \\ 
NGC\,3077      & $ \phn42.13 $             & $14.80\pm1.18 $           &$ 25.11 \pm  2.51 $    &$ \phs\phn14\pm  6 $   &6, 7, 7, 2 \\ 

NGC\,4214      & $ 320\pm 2 $             & $17.87\pm0.08 $           &$ 29.04 \pm 0.11 $        &$\phs291\pm 3 $       & 8, 9, 9, 2 \\ 

NGC\,4449      & $ \phn41.72 $             & $36.00\pm3.00 $           &$ 73.00 \pm 20.00 $        &$ \phs202\pm  7 $     &10, 4, 4, 2 \\ 
NGC\,5253      &$ \phn34.80\pm 5.10 $  & $31.24\pm3.44 $           &$ 29.78 \pm  3.28 $    &$ \phs403\pm 13 $         &11, 4, 4, 2 \\ 
He\,2-10      &$ \phn17.09\pm 1.04 $  & $24.08\pm2.65 $           &$ 26.40 \pm  2.90 $    &$ \phs874\pm 9 $      &12, 4, 4, 2  \\
\hline

\end{tabular}
\label{tab:sample_props_other}
\end{minipage}

\flushleft {\sc References:} (1) \citet{vze98} (\hi-A in their notation); (2) \citet{bot90}; (3) \citet{tul81}; (4) \citet{bei88}; 
(5) \citet{isr88}; (6) integrated over optical counterpart \citet{wal02}; (7) \citet{hun86b}; (8) \citet{huc85}; 
(9) \citet{soi89}; (10) based on the HI image of \citet{hun99}, including the prominent \hi\ ring around the optical counterpart; 
(11) \citet{huc89}; (12) \citet{sau97}

\end{table*}

In this section we compile complementary data on the galaxies in our
sample taken from the literature. The sample consists of eight dwarf
starburst galaxies: I\,Zw\,18, VII\,Zw\,403, NGC\,1569, NGC\,3077,
NGC\,4214, NGC\,4449, NGC\,5253, and He\,2--10 (see Paper\,I). We took
care, as much as possible, to present the data in a homogeneous
manner. We checked the literature for optical, far-infrared (FIR) and
\hi\ properties of our sample. The data were selected to only include
directly observed parameters such as, e.g., fluxes. In
Table\,\ref{tab:sample_props_opt}, we list distances and optical data;
the distance ($D$) to the objects is listed in Column 2. Columns 3 and
4 give the total blue magnitude corrected for Galactic extinction and
inclination ($m_{\rm B}^{0}$) as well as the $K$-band magnitude. The
diameter at the 25\,mag~arcsec$^{-2}$ isophote ($d_{\rm 25}$) is
tabulated in Column 5 and the H$\alpha$ fluxes ($F_{\rm H\alpha}$) in
Column 6. Finally, the oxygen abundance ($12+\log[O/H]$) is given in
Column 7. The last column presents references to the literature that
was used to retrieve the information listed in columns 2 to 7. The
Columns in Table\,\ref{tab:sample_props_other} show the following:
Column 2: \hi\ flux ($F_{\rm HI}$), Columns 3 and 4: FIR $60\mu$ and
$100\mu$ flux densities ($S_{\rm 60\mu}$ and $S_{\rm 100\mu}$), Column
5: the heliocentric radial systemic velocities ($v_{\rm hel}$) as
derived from radio measurements. Again, the last column lists the
references for the data presented in Columns 2 to 5. Note that the
large scale tidal tails around I\,Zw\,18, NGC\,3077 and NGC\,4449 were
not included in the analysis of the \hi\ data.

The derived properties of the galaxy sample were compiled from the
direct measurements given in Tables\,\ref{tab:sample_props_opt} and
\ref{tab:sample_props_other} and are shown in
Table\,\ref{tab:sample_derived}. Whenever no errors were available, we
assumed a conservative 30 per cent uncertainty for the observed
parameters. The absolute magnitudes $M_{\rm B}$ (Column 2) were
derived from the equation $m_{\rm B}^{0}-M_{\rm B}=5\log D - 5 $,
where the distance $D$ is given in parsec. The errors were based on
those given for $m_{\rm B}^{0}$. Blue luminosities $L_{\rm B}$ (Column
3) were calculated via $L_{\rm B}=10^{-0.4(M_{\rm B}-M_{\rm
B_{\odot}})}\;L_{\rm B_{\odot}}$; following \citet{lan92}, we assumed
an absolute blue solar magnitude of $M_{\rm
B_{\odot}}=5.50$\,mag. $K$-band luminosities (Column 4) were derived
in an analogous fashion but with $M_{\rm K_{\odot}}=3.33$\,mag
\citep{cox00}. Furthermore, we used the solar oxygen abundance
$12+\log(O/H)=8.9$ determined by
\citet{lam78} to express the metallicity $Z$ in solar units (Column
5).

\ha\ luminosities (Column 6) were derived from the corresponding
fluxes by multiplying with the geometrical factor $4\pi\,D^{2}$. We
used the conversion established by \citet{ken83}, to obtain the \ha\
star formation rates ($SFR_{\rm H\alpha}$, Column 7): $SFR_{\rm
H\alpha} = L_{\rm H\alpha}/ (1.12\times 10^{41} {\rm{erg~s^{-1}}})
{\rm \,M_{\odot}~yr^{-1}}$.

\begin{table*}
\begin{minipage}{170mm}
\caption
{Derived quantities of the galaxy sample.}
\begin{tabular}{@{}llllllc@{}}
\hline

\multicolumn{1}{c}{Galaxy}&\multicolumn{1}{c}{$M_{\rm B}$}&\multicolumn{1}{c}{$L_{\rm B}$}&\multicolumn{1}{c}{$L_{\rm K}$}&\multicolumn{1}{c}{$Z$}&\multicolumn{1}{c}{$L_{\rm H\alpha}$}&\multicolumn{1}{c}{$SFR_{\rm H\alpha}$}\\

\multicolumn{1}{c}{}&\multicolumn{1}{c}{[mag]}&\multicolumn{1}{c}{[$10^{8} L_{\rm B_\odot}$]}&\multicolumn{1}{c}{[$10^{8} L_{\rm K_\odot}$]}&\multicolumn{1}{c}{[solar]}&\multicolumn{1}{c}{[$10^{38}$\,erg~s$^{-1}$]}&\multicolumn{1}{c}{[M$_{\odot}$\,yr$^{-1}$]}\\

\multicolumn{1}{c}{(1)}&\multicolumn{1}{c}{(2)}&\multicolumn{1}{c}{(3)}&\multicolumn{1}{c}{(4)}&\multicolumn{1}{c}{(5)}&\multicolumn{1}{c}{(6)}&\multicolumn{1}{c}{(7)}\\

\hline

I\,Zw\,18      & $ -14.01\pm0.17 $ & $  \phn0.65\pm  0.10  $ &  \multicolumn{1}{c}{\nodata }&   0.02   &$  61.9\pm      18.6    $&$ 0.06\pm0.02          $\\
VII\,Zw4\,03   & $ -12.53\pm0.3  $ & $  \phn0.16\pm  0.07  $  & \multicolumn{1}{c}{\nodata } &  0.07   &$ 472.5\pm  10.0           $ &$ 0.42\pm 0.01         $\\
NGC\,1569      & $ -18.42\pm0.32 $ & $ 36.98\pm     11.00$ &$\phn7.45\pm0.13$ &  0.21       &$ 479.5\pm143.9                 $&$ 0.43\pm0.13          $\\
NGC\,3077      & $ -17.59\pm0.11 $ & $ 17.22\pm  1.70      $ &$44.06\pm0.81$& 1.00           &$ \phn 58.2\pm 17.5         $&$ 0.05\pm0.02         $\\

NGC\,4214      & $-17.14\pm0.21 $ & $11.38\pm     0.35      $ &$12.36\pm0.58$&0.25       &$147.9\pm44.4                 $&$0.13\pm0.04          $\\
NGC\,4449      & $ -18.31\pm0.62 $ & $ 33.42\pm     20.00      $ &$40.93\pm1.53$&  0.26       &$ 369.4\pm110.8                 $&$ 0.32\pm0.10          $\\
NGC\,5253      & $ -17.82\pm0.40 $ & $ 21.28\pm  8.00      $ &$11.27\pm0.31$&  0.21       &$ 139.4\pm  41.8            $&$ 0.12\pm0.04          $\\
He\,2-10      & $ -17.80\pm0.01 $ & $ 20.89\pm  1.80      $ &$43.65\pm0.81$& 1.07           &$ 339.2\pm101.8                 $&$ 0.30\pm0.09          $\\

\hline

\end{tabular}
\label{tab:sample_derived}
\end{minipage}
\normalsize
\end{table*}

\begin{table*}
\addtocounter{table}{-1}
\begin{minipage}{170mm}
\caption
{-- continued.}
\begin{tabular}{@{}llllllc@{}}
\hline

\multicolumn{1}{c}{Galaxy}&\multicolumn{1}{c}{$M_{\rm HI}$}&\multicolumn{1}{c}{$L_{\rm FIR}$}&\multicolumn{1}{c}{$SFR_{\rm FIR}$}&\multicolumn{1}{c}{$M_{\rm dust}$}&\multicolumn{1}{c}{$t_{\rm life}$}&\multicolumn{1}{c}{$M_{\rm HI}/L_{\rm B}$}\\
\multicolumn{1}{c}{}&\multicolumn{1}{c}{[$10^{8}$\,M$_{\odot}$]}&\multicolumn{1}{c}{[$10^{8} L_{\rm FIR,\odot}$]}&\multicolumn{1}{c}{[$M_{\odot}$\,yr$^{-1}$]}&\multicolumn{1}{c}{[$10^{4}$\,M$_{\odot}$]}&\multicolumn{1}{c}{[Gyr]}&\multicolumn{1}{c}{[$M_{\odot}/L_{\rm B_{\odot}}$]}\\
\multicolumn{1}{c}{}&\multicolumn{1}{c}{(8)}&\multicolumn{1}{c}{(9)}&\multicolumn{1}{c}{(10)}&\multicolumn{1}{c}{(11)}&\multicolumn{1}{c}{(12)}&\multicolumn{1}{c}{(13)}\\

\hline

I\,Zw\,18      &$ 0.4\pm0.01 $& \multicolumn{1}{c}{\nodata} & \multicolumn{1}{c}{\nodata} &$0.4\pm0.2$\footnote{taken from \citet{can02}}&0.7                      &0.62\\ 
VII\,Zw4\,03   &$ 0.7\pm0.2  $&$  0.3\pm0.04           $&$0.005\pm0.001$&$  0.5\pm0.3           $ &0.2                          &4.4\\ 
NGC\,1569      &$ 1.1\pm0.2  $&$  5.9\pm0.7            $&$0.102\pm0.012$&$  2.5\pm0.7          $ &0.3                       &0.03\\
NGC\,3077      &$ 1.3\pm0.4  $&$  5.8\pm0.6            $&$0.100\pm0.010$&$  5.7\pm2.9           $&2.6                      &0.08\\ 

NGC\,4214      &$6.4\pm0.1   $&$4.5\pm1.4                $&$0.078\pm0.024$&$4.0\pm2.0              $&4.9                      &0.6\\

NGC\,4449      &$ 1.5\pm0.5  $&$ 17.9\pm2.8                $&$0.309\pm0.048$&$ 25.7\pm12.9              $&0.5                      &0.04\\
NGC\,5253      &$ 0.9\pm0.1  $&$  8.5\pm0.8            $&$0.147\pm0.014$&$  2.6\pm1.3           $&0.8                      &0.04\\
He\,2-10      &$ 3.3\pm0.2  $&$ 51.9\pm5.2                $&$0.895\pm0.090$&$ 20.6\pm10.3              $&1.1                      &0.16\\

\hline

\end{tabular}
\end{minipage}
\normalsize
\end{table*}

For the conversion of \hi\ flux to the corresponding mass ($M_{\rm
HI}$, given in Column 8 of Table\,\ref{tab:sample_derived}) the
following equation is used: $M_{\rm HI}=2.36\times 10^{5} D^{2} F_{\rm
HI}$ \citep[][with $F_{\rm HI}$ in Jy\,\kms, $M_{\rm HI}$ in
$M_{\odot}$, $D$ in Mpc]{roh00}. The $S_{\rm 60\mu}$ and $S_{\rm
100\mu}$ FIR flux densities were combined in the usual manner to
calculate the FIR luminosity ($L_{\rm FIR}$ in units of $L_{\rm
FIR_{\odot}}$) via $L_{\rm FIR}=3.94\times
10^{5}\,D^{2}\,(2.58\,S_{\rm 60\mu}+S_{\rm 100\mu})\,C$ \citep[][
Column 9]{sol97}. Here, $D$ is the distance in Mpc and $C$ is a colour
correction, which depends on the $S_{\rm 60\mu}/S_{\rm 100\mu}$
ratio. The colour correction varies in the range of $1.5<C<2.1$ and for
simplicity we adopted a common, mean colour correction of $C=1.8$ for
all galaxies. This range for C has been incorporated in the errors
quoted for the fluxes. Using the conversion factor of \citet{ken98}
we derive FIR star formation rates $SFR_{\rm FIR}$ via $SFR_{\rm
FIR}=L_{\rm FIR}/(5.8\times10^{9}\,L_{\rm FIR,\sun}) {\rm
M_{\odot}\,yr^{-1}}$ (Column 10). Furthermore, masses of the cool
dust ($M_{\rm dust}$ in units of $M_{\odot}$; Column 11) were
estimated from the FIR $60\mu$ and $100\mu$ flux densities (in Jy) by
using the equation given in
\citet{huc95}: $M_{\rm dust}=4.78\,S_{\rm
100\mu}\,D^{2}\,(exp[2.94\,(S_{\rm 100\mu}/S_{\rm
60\mu})^{0.4}]-1)$. This equation, however, was derived for a simple
dust model assuming a single temperature of $\sim 40$\,K. As this
needs not to be the case for the galaxies in our sample, we generally
adopt a higher uncertainty of 50 per cent. An estimate for the
lifetimes of the starbursts ($t_{\rm life}=M_{\rm HI}/SFR_{H\alpha}$)
is listed in Column 12. Column 13 lists the gas-to-light $M_{\rm
HI}/L_{\rm B}$ ratios.


\section{The X-ray morphology of the starburst dwarf galaxies}
\label{sec:morph}
\begin{figure*}
\centering
\includegraphics[width=18cm]{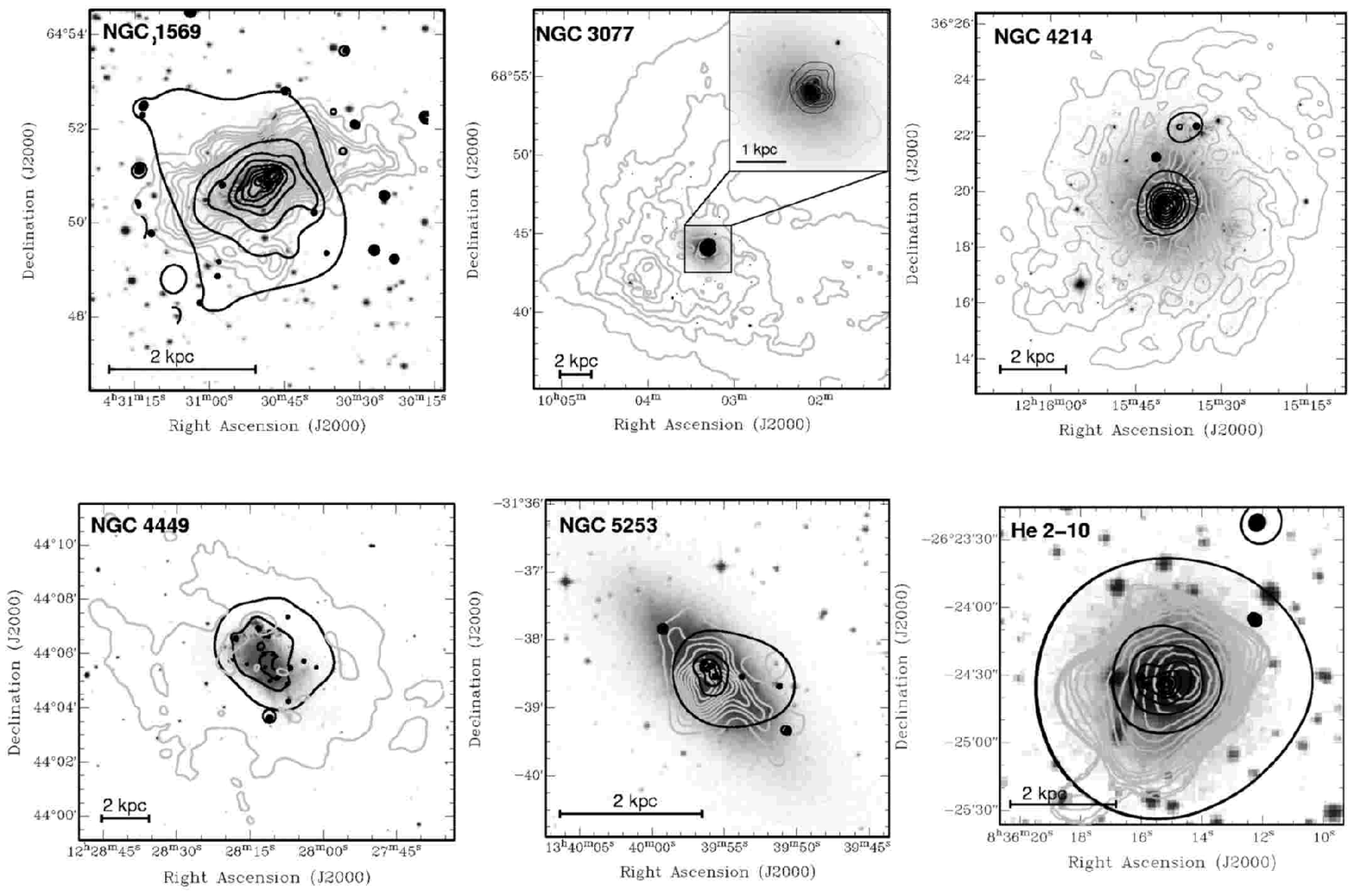}
\caption
{Optical images of dwarf galaxies with detections of diffuse X-rays.
The {\bf black contours} show the adaptively smoothed X-ray
emission. {\bf Grey contours}: \hi\ emission; the contours start at
$10^{20}$\,\cden\ and are spaced by $5\times 10^{20}$\,\cden\ for all
but He\,2-10 (images taken from \citealt{mue03}, \citealt{wal02},
\citealt{hun99}, and \citealt{kob95}). Note the blow-up in the upper 
right corner of the central region of NGC\,3077. The \hi\ contours of
He\,2-10 \citep[taken from][]{kob95b} are not equally spaced; the
outermost contour is at about $10^{20}$\,\cden\ which is 2 per cent of
the peak intensity and the innermost contour is at a level of 95 per
cent (optical images taken from DSS [NGC\,1569], \citealt{wal02}
[NGC\,3077, $R$-band], \citealt{wal01} [NGC\,4214, $B$-band],
\citealt{frei96} [NGC\,4449, $R$-band], \citealt{lau89}[NGC\,5253,
He\,2--10, $R$-band]).}
\label{fig:HI_olays}
\end{figure*}

\begin{figure*}
\centering
\includegraphics[width=18cm]{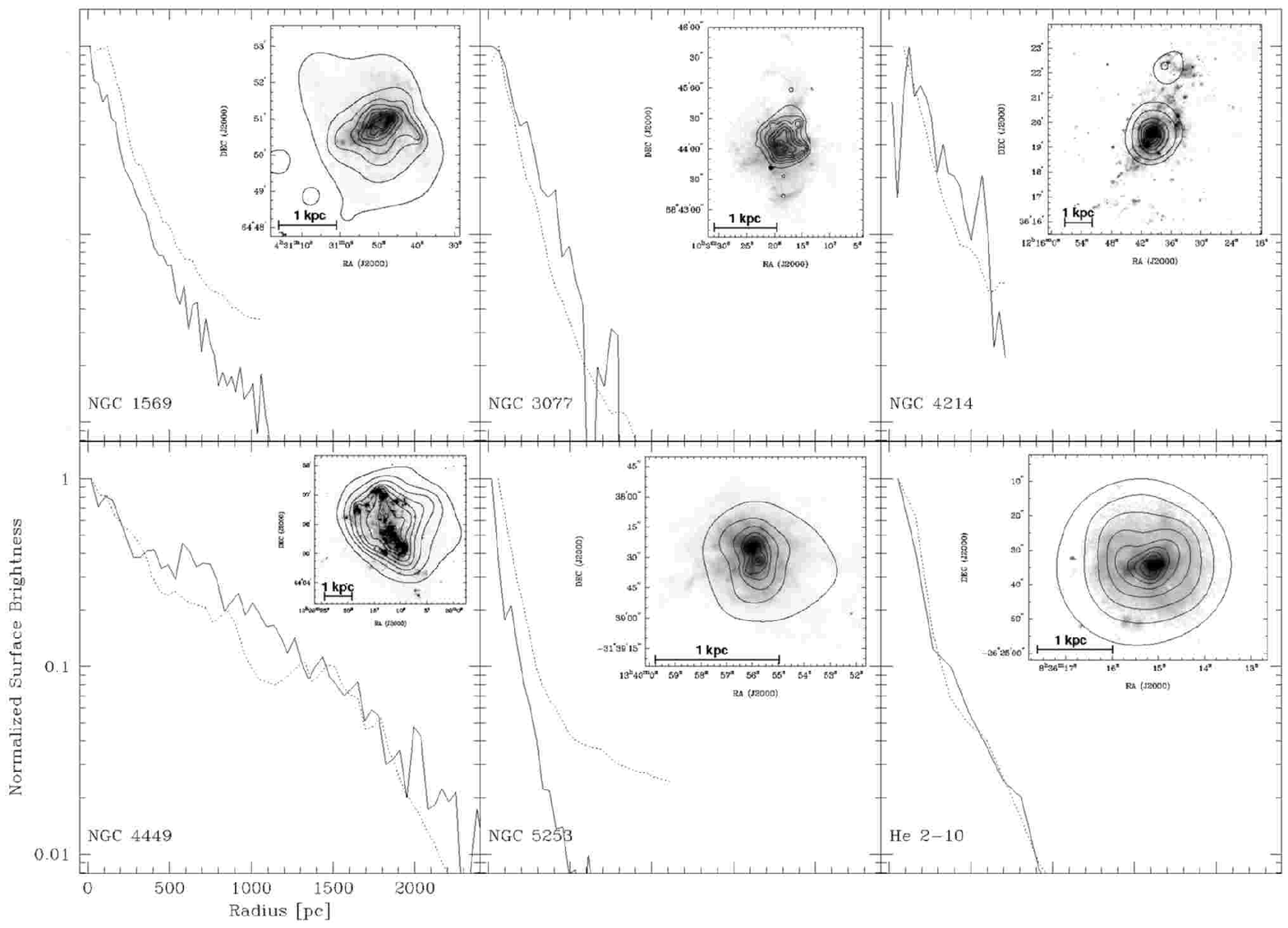}
\caption
{Azimuthally averaged \ha\ ({\bf dotted lines}; excluding bright
compact \hii\ regions) and X-ray ({\bf solid lines}) surface
brightness profiles. All profiles are normalised to their maximum
value and are on the same scale. Note the strong correlation between
the X-ray and
\ha\ profiles. The inserts are overlays of adaptively smoothed diffuse
X-ray emission on the \ha\ images (all logarithmically scaled; see
Paper\,I for details).}
\label{fig:profileboth}
\end{figure*}

Of the eight dwarf starburst galaxies we detect diffuse X-ray emission
in all but two objects (detected in NGC\,1569, NGC\,3077, NGC\,4214,
NGC\,4449, NGC\,5253, and He\,2--10) which can be attributed to hot,
coronal gas stored in galactic winds (Paper\,I). The development of
such a wind is expected to depend on the morphology of the cooler and
denser gas within its host galaxy. In Fig.\,\ref{fig:HI_olays} a
comparison of the optical, \hi\, and X-ray morphologies is made. The
first \hi\ column density contour corresponds to $10^{20}$\,\cden\ and
the outermost X-ray contour is at about the 3--4$\sigma$ detection
limit. The X-ray emission of NGC\,1569 extends well beyond its optical
and \hi\ counterparts perpendicular to the disc. The situation is
different for the other five galaxies: The X-ray, \ha\ (see
Fig.\,\ref{fig:profileboth}), and \hi\ emission of He\,2-10 are all
fairly circularly shaped and have about the same size. Hot gas is
almost entirely bound to the central region of NGC\,4214 (note that
the X-ray emission to the north has low significance and might be an
artefact from the adaptive smoothing process). The optical and X-ray
emission of NGC\,3077 and NGC\,4449 is surrounded by an extended \hi\
envelope which can be ascribed to tidal debris due to a current
interaction \citep[see, e.g., ][]{yun94,hun98}. The \hi\ and X-ray
emission of NGC\,5253 have about the same size, and both are embedded
within a larger optical counterpart. Judged from these 2-dimensional
images, the hot gas in NGC\,3077, NGC\,4214, and NGC\,4449 is still
contained by the cooler, neutral medium. It might be expanding more
freely in NGC\,1569, NGC\,5253, and He\,2-10. The influence of the
neutral ISM might also be obvious from the asymmetric shape of the hot
outflows which is apparent for all galaxies but He\,2-10 and
NGC\,4214. In detail, NGC\,1569 shows outflow features toward the
northern and southern direction. In addition, some of its hot gas
seems to accumulate in a region which is half way between the gaseous
disc and its fainter, southern extension. The well-known arm toward a
westerly direction of NGC\,1569 is visible in optical, \ha\, and X-ray
images (see Fig.\,\ref{fig:profileboth} and Paper\,I). The X-ray
emission of NGC\,3077 is more extended toward the north than toward
the south
\citep[see also Fig.\,\ref{fig:profileboth} and][hereafter OMW03]{ott03}. 
Finally, NGC\,4449 and NGC\,5253 show an extension in X-rays toward a
westerly direction. The hot gas seems always to be elongated along a
direction which corresponds to that of least resistance, along the
steepest gradient in \hi, as can be expected from the hydrodynamics of
a hot, tenuous plasma which interacts with a neutral, denser
environment. In absolute terms, the \hi\ gradients of NGC\,1569,
NGC\,5253, and He\,2-10 are much steeper than the extended \hi\
envelopes of NGC\,3077, NGC\,4449, and NGC\,4214. Note, however, that
this analysis is based on the 2-dimensional morphologies. In fact, an
extension of the hot gas along the line of sight is more than likely
but difficult to determine along with the 3-dimensional morphology of
the \hi\ component.

In Fig.\,\ref{fig:profileboth} we compare the \ha\ and diffuse X-ray
morphologies directly by overlaying the corresponding maps. In the
same figure we also compare azimuthally averaged surface brightness
profiles which have been normalised to their peak intensities. Bright
compact \hii\ regions are excluded for this analysis. In general, the
two different profiles are very similar which is also obvious from the
scale lengths of the two profiles listed in
Table\,\ref{tab:profiles_fit}. \citet{mar97} shows that while
photoionisation is the main process for \ha\ emission in the disc of
starburst galaxies, the emission emerging from the halo is mainly due
to shocks (see Sect.\,\ref{sec:intro}). Leaking photons from \hii\
regions, forward shocks emitting \ha\ radiation, and reverse shocks
which thermalise the hot gas to X-ray emitting temperatures are all
the result of the impact of massive stars on the ISM. Since the
starburst are not as much confined to the nuclei of the galaxies as,
e.g., in the massive starburst galaxies M\,82 and NGC\,253, this may
explain the similarity of the
\ha\ and X-ray surface brightness profiles on global scales in the
starburst galaxies (see also Sect.\,\ref{sec:correl}). Due to the
different locations of the forward and reverse shocks, however, the
hot gas and the \ha\ emission do not trace each other exactly on local
scales. E.g., the expanding
\ha\ shells in NGC\,3077 are filled with hot gas (see OMW03), and the
prominent \ha\ arm in NGC\,1569 is somewhat displaced from a similar
feature seen in X-rays \citep[see Fig.\,\ref{fig:profileboth}
and][]{mar02}.


\begin{table}
\centering
\caption
{Scale lengths (in pc) of exponential functions fitted to azimuthally
averaged X-ray (Total band, $h_{\rm X (Total)}$) and \ha\ surface
brightness profiles ($h_{\rm H\alpha}$).}
\begin{tabular}{@{}lcc@{}}
\hline
Galaxy&$h_{\rm X (Total)}$  & $h_{\rm H\alpha}$\\

\hline

NGC\,1569 & $208\pm  \phn6$ & $185\pm \phn8$ \\ 
NGC\,3077 & $186\pm 10$     & $140 \pm \phn4$ \\
NGC\,4214 & $145\pm 12$     & $127\pm \phn3$ \\
NGC\,4449 & $579\pm 15$     & $524\pm 19$ \\
NGC\,5253 & $109\pm \phn9$  & $109\pm \phn3$ \\
He2-10   & $186\pm 15$     & $227\pm \phn9$\\

\hline
\end{tabular}
\label{tab:profiles_fit}
\end{table}


\section{X-ray point source populations}
\label{sec:point_pop}

\subsection{Distribution of X-ray point sources}

\begin{figure*}
\centering
\includegraphics[width=18cm]{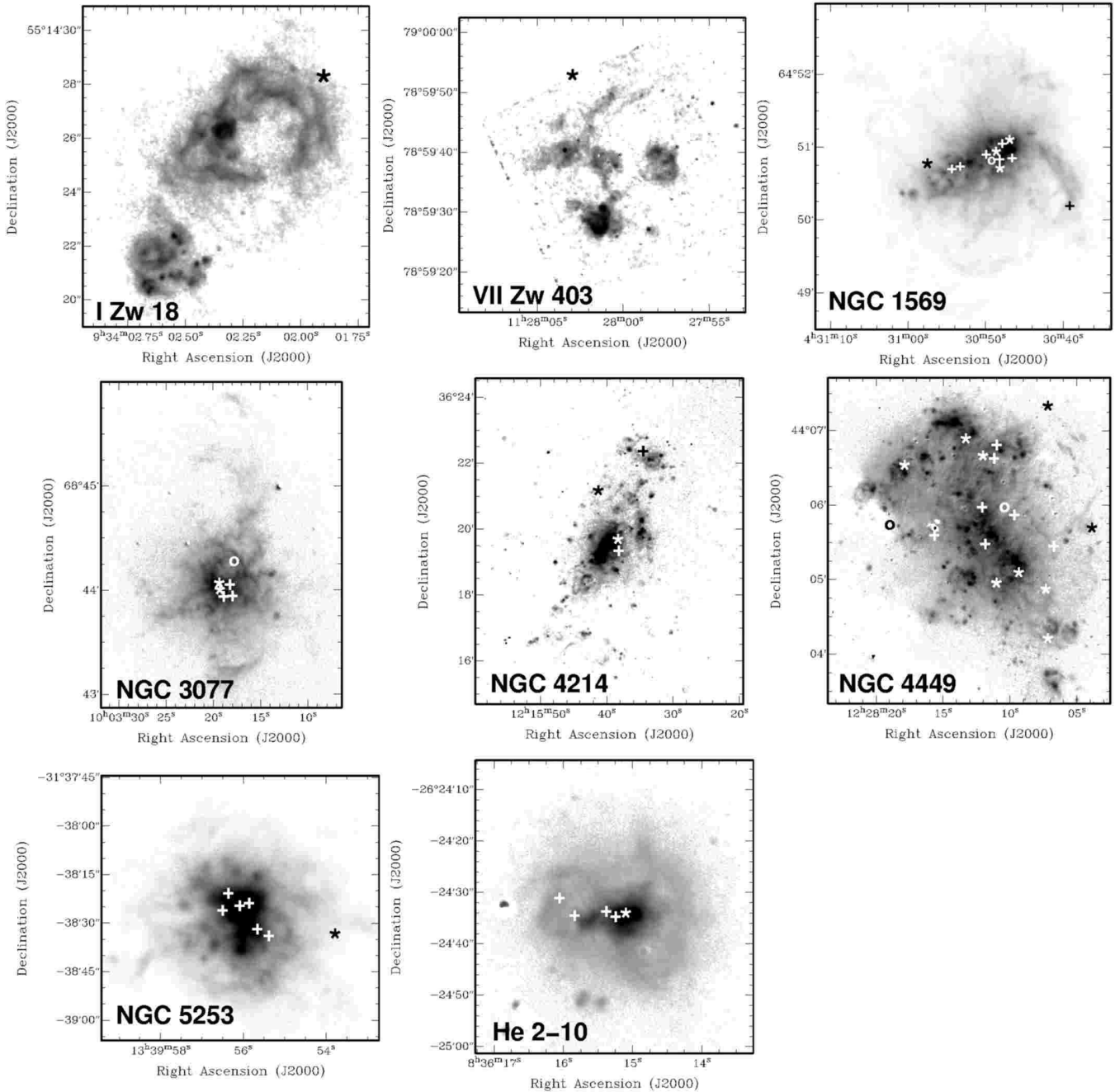}
\caption
{\ha\ images of the galaxies in the sample (logarithmic scale). The
  location of X-ray point sources are overplotted and marked
  according to their spectral type: {\bf Star:} power law, {\bf Cross:}
  thermal plasma, and {\bf Circle:} black body spectrum. }
\label{fig:ha_point}
\end{figure*}

X-ray point sources in the galaxy sample are detected mainly close to
the centres of the galaxies. The only exception is NGC\,4449 where the
sources are more uniformly scattered across the entire optical body
(see Fig.\,\ref{fig:ha_point}). The reason might be that NGC\,4449 has
a different star formation history as compared to the other
galaxies. This is exemplified by the \ha\ emission in NGC\,4449 which
is more uniformly distributed over its optical body whereas again the
other galaxies in the sample show a more concentrated surface
brightness (Fig.\,\ref{fig:profileboth}; see also the \ha\ and X-ray
scale lengths which are for NGC\,4449 a factor of 2.5 larger than for
any other galaxy: Table\,\ref{tab:profiles_fit}).

In general, the brightest population of extragalactic X-ray point
sources are AGNs, X-ray binaries (XRBs), SNRs, and supersoft
sources. Except for AGNs, all of these sources need massive
(progenitor) stars for their X-ray emission. For this reason they are
expected to concentrate near star forming regions or young stellar
clusters. In Fig.\,\ref{fig:ha_point} we overlay the location of the
X-ray point sources on the \ha\ emission and indicate the type of
model which best fits their spectra. The bulk of the point sources are
detected in the vicinity of bright \hii\ regions or on the rims of
supergiant shells. Only a few sources cannot be related to individual
\ha\ features. There is no obvious trend for a source type with
galactocentric distance (see Paper\,I for details). Note that a
certain fraction of point sources might also be unrelated foreground
or background sources which are simply projected along the same line
of sight as the galaxies under consideration.

\begin{figure*}
\includegraphics[width=15cm]{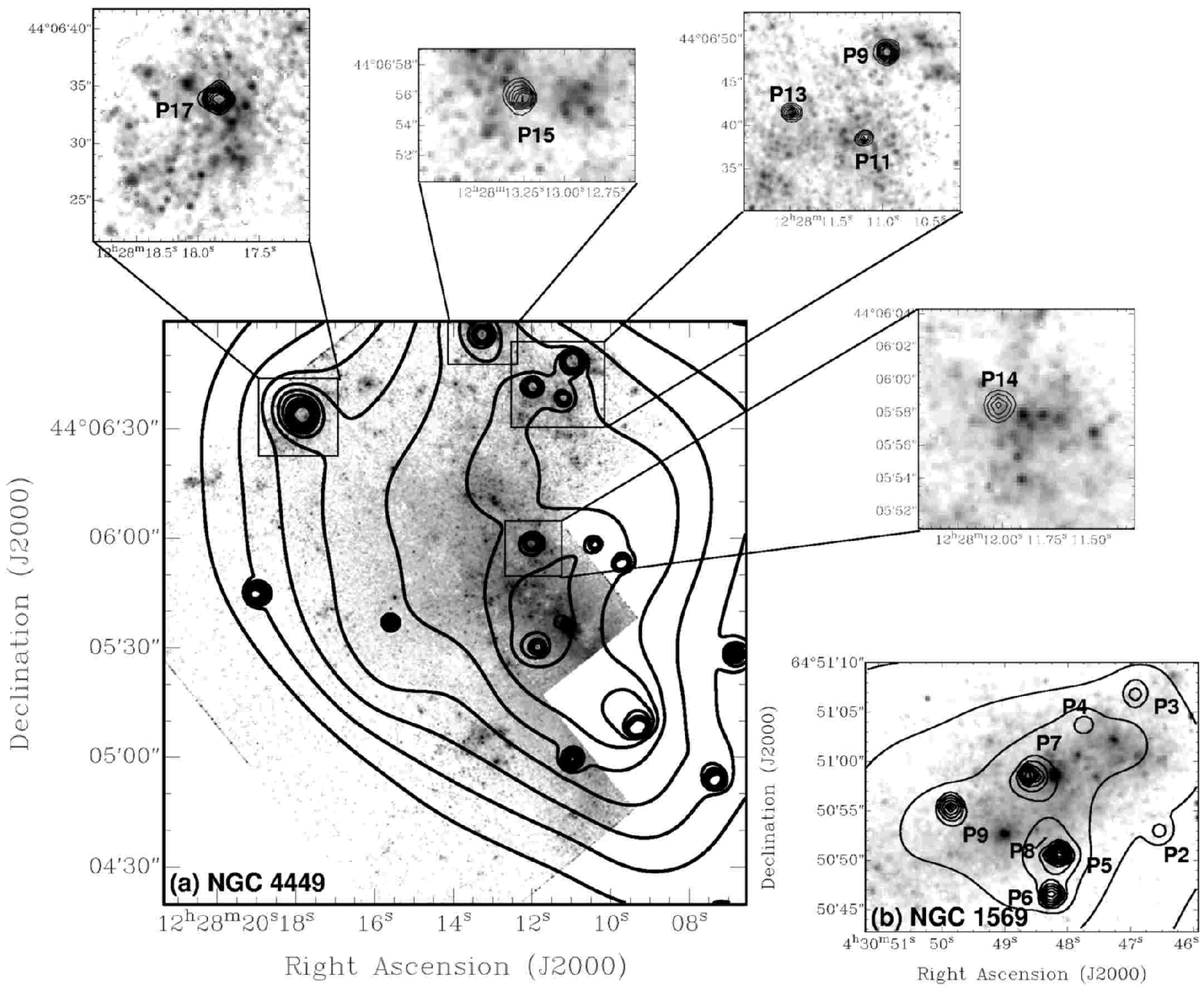}
\caption
{{\it HST}/WFPC2 images of NGC\,4449 {\bf (a)} and NGC\,1569 {\bf (b)}
  in the F336W (Johnson $U$-band). Overlaid are contours of the {\it
    Chandra} X-ray observations. In some areas X-ray point sources are
  close to bright young stellar clusters. In the case of NGC\,4449
  these regions are shown as blow-ups of the larger figure.}
\label{fig:n4449_U}
\end{figure*}

Young, massive stars are very bright at ultraviolet wavelengths.
Consequently, if X-ray point sources are related to star forming
regions, they should be spatially correlated with UV bright stellar
clusters (note, however, that variable internal extinction can make a
correlation more uncertain). F336W (corresponding to Johnson $U$-band)
images were available in the {\it HST}/WFPC2 archive for NGC\,4449 and
NGC\,1569 and in Fig.\,\ref{fig:n4449_U} we show X-ray contours with
labelled point sources overlaid on these images (for the labelling of
the sources see Paper\,I). Almost all of the central X-ray point
sources in NGC\,1569 and most of the point sources in NGC\,4449 which
are covered by the WFPC2 field of view are located very close to UV
bright stellar clusters but often slightly displaced. Those sources
are likely associated with recent star formation events. There is no
relation, though, between the location of these sources and the shape
of their X-ray spectrum, be it represented by power law (PL), thermal
plasma (TP), or black body (BB) shapes (see also
Fig.\,\ref{fig:ha_point} and Paper\,I). This is in conflict with the
models by \citet{kar04} which imply that the slightly displaced
sources are all XRBs.

\subsection{X-ray point sources as a measure for star formation}
\label{sec:point_sfr}

\begin{figure*}
\centering
\includegraphics[width=18cm]{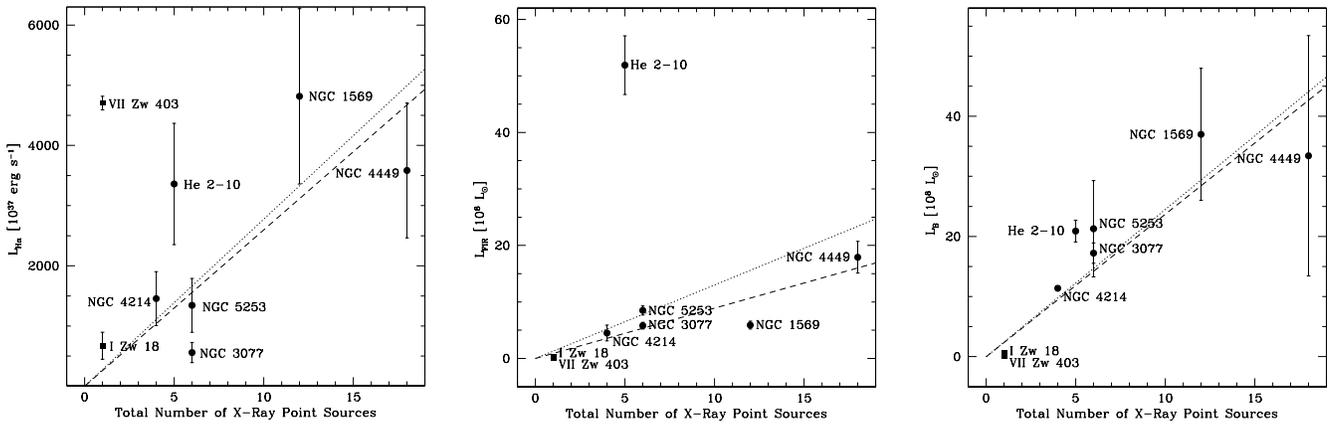}
\caption
{Star formation tracers \ha, FIR, and blue luminosity plotted against
the number of X-ray point sources. Fits including (excluding)
He\,2-10 are displayed by {\bf dashed} ({\bf dotted}) lines. }
\label{fig:correl_point}
\end{figure*}

The proximity of the X-ray sources and the \hii\ regions/young stellar
clusters suggests that the number of X-ray point sources might be a
tracer for SF in general. Such a correlation was suggested by
\citet{gri03} for high-mass XRBs. In Fig.\,\ref{fig:correl_point} we
plot the number of X-ray point sources as a function of different SF
tracers: the \ha, the FIR, and the blue ($B$-band) luminosity. Trends
are apparent for all three SF tracers. The correlation is best for the
FIR luminosity and the blue luminosity. There is a larger scatter in
the case of the \ha\ luminosities. Least square fits to the data
result in the following equations ($n$: number of X-ray point
sources). $SFR_{\rm H\alpha}/$[\msun\,yr$^{-1}$]$=(0.025\pm0.008)\,
n$, $SFR_{\rm FIR}/$[\msun\,yr$^{-1}$]$=(0.022\pm0.013)\,n$, and
$L_{\rm B}/$[$10^{8}$\,L$_{B\sun}$]$=(2.45\pm0.27)\,n$. He\,2-10 is
about three times more distant than the majority of the
sample. Therefore, the detection limit is higher and point sources
with a low X-ray flux will fall below this limit (for detection
limits, see Paper\,I). For this reason we fitted the same relations
excluding this galaxy, which yields $SFR_{\rm
H\alpha}/$[\msun\,yr$^{-1}$]$=(0.023\pm0.008)\,n$, $SFR_{\rm
FIR}/$[\msun\,yr$^{-1}$]$=(0.015\pm0.002)\,n$, and $L_{\rm
B}/$[$10^{8}$\,L$_{B\sun}$]$=(2.37\pm0.26)\,n$. These relations
suggest that the starburst is responsible not only for the hot gas but
also for the X-ray point sources (see also
Fig.\,\ref{fig:correl_starformation}).

\subsection{Point source luminosity function}

\begin{figure*}

\includegraphics[width=18cm]{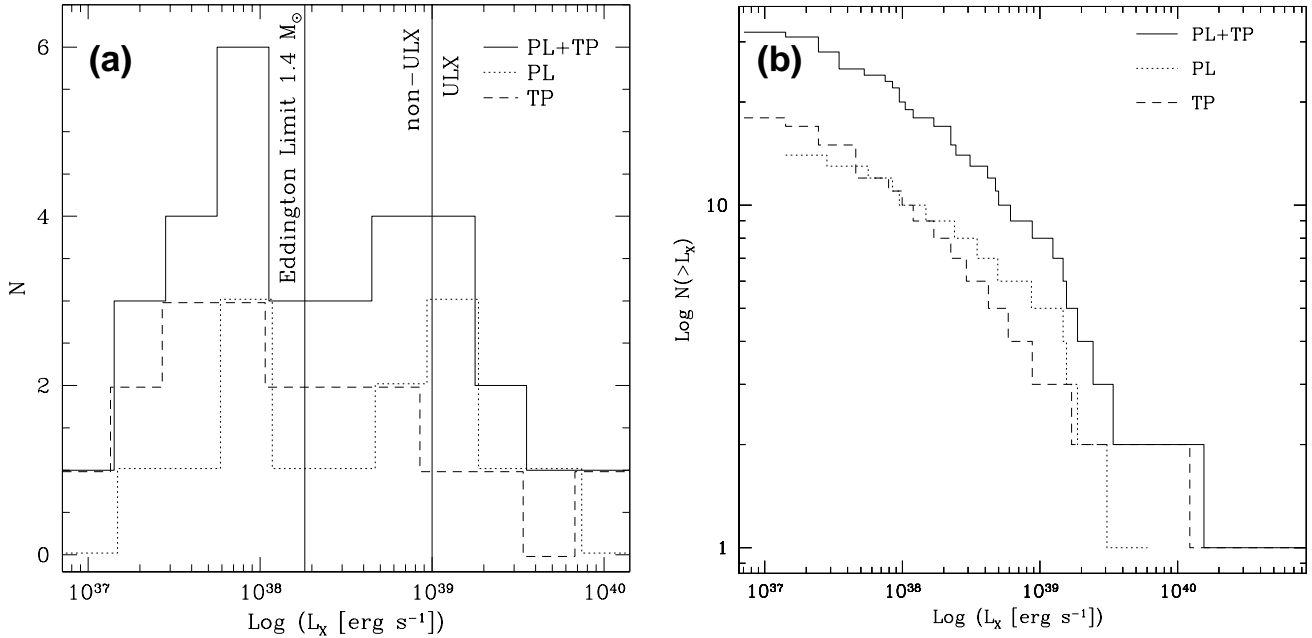}
\caption
{{\bf (a)} Differential luminosity function of the X-ray point sources
within all galaxies in the sample best fit by a MeKaL thermal plasma
({\bf TP}) or power law ({\bf PL}) spectrum as well as the combination
of these two functions. $N$ is the number of point sources and $L_{\rm
X}$ the X-ray luminosity. The {\bf left} vertical line displays the
Eddington limit for a 1.4\,M$_{\sun}$ source and the {\bf right}
vertical line marks the threshold for ultraluminous X-ray sources
({\bf ULXs}). {\bf (b)} Cumulative version of (a).}
\label{fig:logn}
\end{figure*}

A differential as well as a cumulative luminosity function of all the
point sources in the galaxies is shown in Fig.\,\ref{fig:logn}. In
this plot, we mark the location of the Eddington luminosity limit of a
1.4\,\msun\ star (a star at the Chandrasekhar limit), i.e., the X-ray
luminosity at which the photon pressure equals the gravitational pull of
the receptor star. In addition, we plot the luminosity limit for
ultraluminous X-ray sources (ULX, definition: $L_{\rm
X}>10^{39}$\,\lum) which in fact is the Eddington luminosity of a
$\sim 7.7$\,M$_{\sun}$ object, a mass which is often used as a lower
limit for SNe progenitor stars
\citep[e.g.,][]{lei99}. About ten X-ray point sources in our sample
can be classified as ULXs.  These sources -- intermediate-mass black
holes or high-mass XRBs with accretion above the Eddington limit
\citep[for a discussion of ULXs, see, e.g.,][]{kin04} -- are only
found in I\,Zw\,18 (one source), NGC\,4449 (five, including two which
are best fitted by black body spectra), and He\,2-10 (four). These
figures are even larger than the average number of ULXs in massive but
non-starburst galaxies
\citep[$\sim 2$, see][]{col02}. In merging galaxies such as the
Antennae, however, the number of ULXs appears to be substantially
increased \citep{zez02}. Indeed, I\,Zw\,18 and NGC\,4449 are clearly
interacting galaxies \citep{vze98,hun98,the01} and for He\,2-10 a
merger scenario has been proposed in the literature
\citep{kob95b,con94}. However, NGC\,3077 is also interacting but does
not have any ULXs. On the basis of our limited sample it is not
possible to derive any clear dependence between the strength of the
galaxy interaction and the number of ULXs.

A power law fit to the cumulative luminosity distribution for all
sources results in a slope of $\alpha=0.24\pm0.06$ ($N(>L_{\rm
X})\propto L_{\rm X}^{-\alpha}$). This value has been derived using
the data of the differential luminosity distribution following the
method described in \citet{cra70}. Due to our completeness limits,
data bins with luminosities less than $10^{38}$\,\lum\ were discarded
for this fit. Similar fits for the individual PL and TP source
populations lead to very similar slopes of $\alpha=0.23\pm0.08$ (PL)
and $0.25\pm0.08$ (TP). The power law index derived here for the dwarf
starburst galaxies is shallower than those observed in more massive
starbursts such as NGC\,253, M\,82, and the Antennae (indices of
$\sim0.4-0.8$). The power law indices of non-starburst galaxies
($\alpha\sim1.2$) are far steeper than both, the value for our sample
and the power law indices of more massive starburst galaxies
\citep[for a comparison of starburst and
non-starburst cumulative luminosity functions see][]{kil02}.


\section{X-ray properties of the hot coronal gas}

\label{sec:xrayprop}

In Paper\,I we performed one-temperature MeKaL thermal equilibrium
plasma fits \citep{mew85,kaa92,lie95} to the diffuse X-ray emission
within certain regions of the galaxies. In addition to fitting the
total emission due to diffuse gas we also made fits to the emission
from the centre and various regions around it.
Table\,\ref{tab:detect_gasb} shows the main fitting results listing
the absorbing column densities $N_{\rm H}$, the temperature $T$, the
normalisation, absorbed X-ray flux $F_{\rm X}^{\rm abs}$, unabsorbed
X-ray flux $F_{\rm X}$, and unabsorbed X-ray luminosity $L_{\rm X}$.
The fits are used to determine intrinsic parameters of the hot gas:
the electron volume density $n_e$ follows from the normalisation of
the spectra assuming a spherical geometry and that the proton and
electron density are the same over the volume $V$ (see OMW03); other
parameters that can be derived are the equivalent diameter $d_{eq}$ of
the X-ray emitting region ($d_{eq}=\sqrt{4 Area/\pi}$, where $Area$ is
the area of the emission), the mean line of sight $l$, the emission
measure $EM=n_e^{2}l$, the pressure of the hot gas $P/k=2\,n_{e}\,T$
($k$: Boltzmann's constant), its mass $M_{\rm hot}=n_{\rm e} m_{\rm p}
V$ ($m_{\rm p}$: proton mass, $V$: shell volume), the thermal energy
$E_{\rm th}=3 n_{\rm e} V kT$, the cooling time $t_{\rm cool}=E_{\rm
th}/L_{\rm X}$, the mass deposition rate $\dot{M}_{\rm cool}=M_{\rm
hot}/t_{cool}$, and finally the mean thermal velocity per particle
$<v_{\rm hot}>= \sqrt{2 E_{\rm th}/M_{\rm hot}}$. The quantities for
the different galaxies are listed in
Table\,\ref{tab:common_gasparams}. Note that most of these values
depend on the volume filling factor $f_{v}$ and the metallicity
$Z$. For the metallicities we use those derived for \hii\ regions in
the individual objects (for a discussion see Paper\,I). As $f_{v}$ is
a priori unknown, we derive all gas parameters for $f_{v}=1$, any
corrections for lower filling factors must be applied according to the
table header in Table\,\ref{tab:common_gasparams}. High angular
resolution {\it Chandra} observations of more massive starburst
galaxies \citep[see, e.g., ][and references therein]{hec02} and
hydrodynamical simulations \citep[e.g., ][]{str00} reveal that the
volume filling factor falls most likely in the range of 0.1--0.3 for
superwinds. Confined shells, however, such as those in NGC\,3077 (see
OMW03) may have larger filling factors. An observational determination
of the filling factor for dwarf galaxies is not an easy task due to
their lower flux and the unknown source geometries. In addition,
hydrodynamical simulations show that the filling factor can take up
very different values. For the rest of this paper we assume that
$f_{v}$ is about the same for all objects and that therefore the
values may be compared, at least in a relative sense.

\begin{table*}
\begin{minipage}{170mm}
\centering
\caption
{Results of fits of collisional thermal plasma models to the X-ray
spectra applied to the entire galaxy, the centre and outer
regions. This table is a shortened version of the corresponding tables
in Paper\,I (fluxes and luminosities are given within an energy range
of $0.3-8.0$\,keV).}
\begin{tabular}{@{}lccccccc@{}}
\hline

\multicolumn{1}{l}{Region}&\multicolumn{1}{c}{$N_{\rm H}$}&\multicolumn{1}{c}{$T$}&\multicolumn{1}{c}{$Norm^{a}$}&\multicolumn{1}{c}{$F_{\rm X}^{\rm abs}$}&\multicolumn{1}{c}{$F_{X}$}&\multicolumn{1}{c}{$L_{X}$}&\multicolumn{1}{c}{$\chi^{2}_{\rm red}$}\\
\multicolumn{1}{c}{}&\multicolumn{1}{c}{[$10^{21}$\,cm$^{-2}$]}&\multicolumn{1}{c}{[$10^{6}$\,K]}&\multicolumn{1}{c}{[$10^{-5}$]}&\multicolumn{1}{c}{[$10^{-15}$\,erg~s$^{-1}$\,cm$^{-2}$]}&\multicolumn{1}{c}{[$10^{-15}$\,erg~s$^{-1}$\,cm$^{-2}$]}&\multicolumn{1}{c}{[$10^{37}$\,erg~s$^{-1}$]}&\\

\hline

&\multicolumn{7}{c}{NGC\,1569}\\
\hline
Total            & $2.86_{-0.15}^{+0.15}$ & $7.23_{-0.12}^{+0.12}$ & $80.0_{-1.5}^{+1.5}$ & $205_{-10}^{+10}$ & $721_{-15}^{+15}$ & $41.7_{-6.0}^{+6.0}$ & 0.98\\

Centre           & $4.99_{-0.19}^{+0.19}$ & $7.23_{-0.13}^{+0.13}$ & $55.2_{-1.2}^{+1.2}$ & $116_{-7}^{+7}$ & $497_{-13}^{+13}$ & $28.8_{-4.1}^{+4.1}$ & 0.68\\

Outer Regions             &  $9.12_{-0.26}^{+0.27}$ & $3.51_{-0.06}^{+0.06}$ & $125_{-4}^{+5}$ & $58.8_{-5.8}^{+6.2}$ & $809_{-36}^{+36}$ & $46.8_{-6.9}^{+6.9}$ & 0.78\\
\hline
&\multicolumn{7}{c}{NGC\,3077}\\
\hline
Total            & $4.71_{-0.11}^{+0.11}$ & $2.32_{-0.04}^{+0.04}$ & $50.7_{-2.8}^{+2.8}$ & $33.6_{-5.1}^{+5.6}$ & $1062_{-69}^{+71}$ & $165_{-26}^{+26}$ & 0.43\\

Centre           & $0.86_{-0.28}^{+0.32}$ & $6.04_{-0.87}^{+0.72}$ & $0.39_{-0.05}^{+0.05}$ & $6.16_{-1.85}^{+1.95}$& $11.2_{-1.8}^{+1.6}$ & $1.74_{-0.37}^{+0.37}$ & 0.11\\
Outer Regions             & $4.52_{-0.13}^{+0.14}$ & $2.30_{-0.04}^{+0.04}$ & $30.5_{-2.2}^{+2.2}$ & $21.6_{-4.1}^{+4.7}$ & $637_{-53}^{+54}$ & $98.7_{-16.3}^{+16.3}$ & 0.44\\
\hline
&\multicolumn{7}{c}{NGC\,4214}\\
\hline
Total            & $9.87_{-0.52}^{+0.59}$ & $2.17_{-0.06}^{+0.06}$ & $129_{-14}^{+14}$ & $37.9_{-10.6}^{+12.9}$ & $759_{-95}^{+98}$ & $75.9_{-14.5}^{+14.5}$& 0.14\\

Centre           & $10.1_{-1.1}^{+1.4}$ & $2.80_{-0.28}^{+0.27}$ & $22.7_{-5.3}^{+5.3}$ & $12.5_{-6.7}^{+9.8}$ & $155_{-41}^{+44}$ & $15.5_{-4.9}^{+4.9}$ & 0.04\\
Outer Regions  & $2.91_{-0.56}^{+0.67}$ & $2.68_{-0.24}^{+0.21}$ & $12.2_{-1.8}^{+1.8}$ & $25.8_{-9.6}^{+11.3}$ & $82.3_{-15.6}^{+14.9}$ & $8.23_{-1.95}^{+1.95}$ & 0.14\\
\hline
&\multicolumn{7}{c}{NGC\,4449}\\
\hline
Total            & $7.19_{-0.10}^{+0.10}$ & $2.98_{-0.03}^{+0.03}$ & $353_{-7}^{+7}$ & $393_{-20}^{+21}$ & $2481_{-55}^{+55}$ & $452_{-65}^{+65}$ & 1.07\\

Centre           & $7.87_{-0.22}^{+0.23}$ & $3.12_{-0.07}^{+0.07}$ & $71_{-3}^{+3}$ & $75.9_{-8.5}^{+9.2}$ & $510_{-26}^{+26}$ & $92.8_{-14.0}^{+14.0}$ & 0.26\\

Outer Regions             & $4.88_{-0.12}^{+0.13}$ & $3.33_{-0.05}^{+0.05}$ & $111_{-3}^{+3}$ & $225_{-15}^{+15}$ & $821_{-25}^{+26}$ & $149_{-22}^{+22}$ & 0.83\\

\hline
&\multicolumn{7}{c}{NGC\,5253}\\
\hline
Total              & $5.67_{-0.22}^{+0.23}$ & $3.51_{-0.08}^{+0.08}$ & $55.5_{-2.1}^{+2.0}$ & $81.6_{-8.1}^{+8.6}$ & $359_{-17}^{+18}$ & $46.6_{-7.0}^{+7.0}$ & 0.64\\
Centre             & $10.1_{-0.3}^{+0.3}$ & $3.79_{-0.09}^{+0.09}$ & $44.6_{-2.0}^{+2.0}$ & $39.3_{-4.6}^{+5.0}$ & $299_{-17}^{+17}$ & $38.9_{-5.9}^{+5.9}$ & 0.23\\

Outer Regions               &$7.58_{-0.75}^{+0.88}$ & $3.21_{-0.19}^{+0.23}$ & $25.6_{-3.4}^{+3.4}$ & $24.6_{-7.8}^{+10.1}$ & $158_{-25}^{+27}$ & $20.6_{-4.6}^{+4.6}$ & 0.49\\
\hline
&\multicolumn{7}{c}{He\,2-10}\\
\hline
Total              & $4.44_{-0.29}^{+0.35}$ & $2.82_{-0.14}^{+0.14}$ & $93.6_{-13.0}^{+13.0}$ & $103_{-36}^{+46}$ & $2101_{-313}^{+319}$ & $2038_{-423}^{+423}$ & 0.32\\
Centre             &$1.23_{-0.20}^{+0.22}$ & $7.60_{-0.44}^{+0.44}$ & $3.38_{-0.27}^{+0.27}$ & $41.9_{-6.5}^{+6.9}$ & $95.6_{-6.4}^{+6.3}$ & $92.7_{-14.5}^{+14.5}$ & 0.09\\
Outer Regions               & $2.82_{-0.14}^{+0.15}$ & $3.21_{-0.10}^{+0.11}$ & $17.5_{-1.2}^{+1.2}$ & $50.4_{-9.1}^{+9.9}$ & $410_{-31}^{+32}$ & $397_{-64}^{+64}$ & 0.22\\
\hline
\end{tabular}
\footnotetext{$^{a}$ The normalisation is given by $10^{-14} (4\pi D^{2})^{-1} \int n_{\rm
e} n_{\rm H} dV$ with $D$: distance in cm, $n_{\rm
e}, n_{\rm H}$: electron and proton densities in \vden, $V$ volume in cm$^{3}$.}
\label{tab:detect_gasb}
\end{minipage}
\end{table*}

\begin{table*}
\begin{minipage}{170mm}
\centering
\caption
{Derived parameters of the hot gas. All values were determined for a
filling factor $f_{v}=1$ and metallicities $Z_{\rm HII}$ as derived for the
\hii\ regions in the objects (see
Table\,\ref{tab:sample_derived}). The values can be converted to lower
filling factors and other metallicities $Z$ (via $\xi=Z/Z_{\rm HII}$) of
the X-ray emitting gas as given in the head of the table.}

\begin{tabular}{@{}lccccc@{}}
\hline

\multicolumn{1}{l}{Region}&\multicolumn{1}{c}{$d_{eq}$}&\multicolumn{1}{c}{$l$}&\multicolumn{1}{c}{$n_e$}&\multicolumn{1}{c}{$EM$}&\multicolumn{1}{c}{$P/k$}\\

\multicolumn{1}{l}{}&\multicolumn{1}{c}{}&\multicolumn{1}{c}{}&\multicolumn{1}{c}{$(\times [f_{v}\xi]^{-0.5})$}&\multicolumn{1}{c}{$(\times [f_{v}\xi]^{-1})$}&\multicolumn{1}{c}{$(\times [f_{v}\xi]^{-0.5})$}\\

\multicolumn{1}{l}{}&\multicolumn{1}{c}{[pc]}&\multicolumn{1}{c}{[pc]}&\multicolumn{1}{c}{[$10^{-3}$\,cm$^{-3}$]}&\multicolumn{1}{c}{[cm$^{-6}$\,pc]}&\multicolumn{1}{c}{[10$^{5}$\,K\,cm$^{-3}$]}\\

\hline
\multicolumn{6}{c}{NGC\,1569}\\
\hline

Total & $2100\pm 210$ & $1400\pm140$ & $18\pm2$ & $0.45\pm0.05$ & $2.61\pm0.27$ \\
Centre & $820\pm82$ & $540\pm54$ & $61\pm6$ & $2.03\pm0.20$ & $8.87\pm0.91$ \\
Outer Regions &  $1580\pm 158$ & $1060\pm106$ & $35\pm4$ & $1.26\pm0.13$ & $2.42\pm0.25$ \\

\hline
\multicolumn{6}{c}{NGC\,3077}\\
\hline
Total & $1820\pm182$ &  $850\pm85$ & $29\pm3$ & $0.72\pm0.07$ & $1.35\pm0.14$ \\
Centre & $300\pm30$ & $200\pm20$ & $38\pm5$ & $0.29\pm0.03$ & $4.61\pm0.86$ \\
Outer Regions & $1360\pm136$ & $900\pm90$ & $35\pm4$ & $1.10\pm0.11$ & $1.61\pm0.17$ \\

\hline
\multicolumn{6}{c}{NGC\,4214}\\
\hline

Total  & $1270\pm127$ & $850\pm85$ & $64\pm7$ & $3.50\pm0.35$ & $2.78\pm0.33$ \\
Centre & $410\pm41$ & $270\pm27$ & $147\pm23$ & $5.81\pm0.58$ & $8.21\pm1.51$ \\
Outer Regions   & $910\pm91$ & $600\pm60$ & $33\pm4$ & $0.63\pm0.06$ & $1.74\pm0.27$ \\

\hline
\multicolumn{6}{c}{NGC\,4449}\\
\hline

Total  & $4120\pm412$ & $2740\pm274$ & $24\pm2$ & $1.63\pm0.16$ & $1.46\pm0.15$ \\
Centre & $1000\pm100$ & $660\pm66$ & $92\pm9$ & $5.54\pm0.55$ & $5.71\pm0.60$ \\
Outer Regions   & $2820\pm282$ & $1880\pm188$ & $24\pm2$ & $1.10\pm0.11$ & $1.61\pm0.16$ \\

\hline
\multicolumn{6}{c}{NGC\,5253}\\
\hline

Total  &$2340\pm234$ & $1560\pm156$ & $19\pm2$ & $0.57\pm0.06$ & $1.34\pm0.14$ \\
Centre &$840\pm84$ & $560\pm56$ & $80\pm8$ & $3.56\pm0.36$ & $6.05\pm0.64$ \\
Outer Regions   &$1740\pm174$ & $1160\pm116$ & $20\pm2$ & $0.48\pm0.05$ & $1.30\pm0.18$ \\

\hline
\multicolumn{6}{c}{He\,2-10}\\
\hline

Total  & $3180\pm318$ & $2120\pm212$ & $43\pm5$ & $3.88\pm0.39$ & $2.41\pm0.32$ \\
Centre & $500\pm50$ & $340\pm34$ & $130\pm14$ & $5.78\pm0.58$ & $19.83\pm2.42$ \\
Outer Regions  & $1960\pm196$ & $1300\pm130$ & $38\pm4$ & $1.90\pm0.19$ & $2.45\pm0.27$ \\

\hline
\end{tabular}
\label{tab:common_gasparams}
\end{minipage}
\end{table*}


\begin{table*}
\addtocounter{table}{-1}
\begin{minipage}{170mm}
\centering
\caption{--- continued.}
\begin{tabular}{@{}lccccc@{}}
\hline

\multicolumn{1}{l}{Region}&\multicolumn{1}{c}{$M_{hot}$}&\multicolumn{1}{c}{$E_{th}$}&\multicolumn{1}{c}{$t_{cool}$}&\multicolumn{1}{c}{$\dot{M}_{cool}$}&\multicolumn{1}{c}{$<v_{hot}>$}\\

\multicolumn{1}{l}{}&\multicolumn{1}{c}{$(\times f_{v}^{0.5}\xi^{-0.5})$}&\multicolumn{1}{c}{$(\times f_{v}^{0.5}\xi^{-0.5})$}&\multicolumn{1}{c}{$(\times f_{v}^{0.5}\xi^{-0.5})$}&\multicolumn{1}{c}{}&\multicolumn{1}{c}{}\\

\multicolumn{1}{l}{}&\multicolumn{1}{c}{[$10^{4}$\,M$_{\odot}$]}&\multicolumn{1}{c}{[$10^{52}$\,erg]}&\multicolumn{1}{c}{[Myr]}&\multicolumn{1}{c}{[$10^{-3}$\,M$_{\odot}$\,yr$^{-1}$]}&\multicolumn{1}{c}{[km\,s$^{-1}$]}
\\

\hline
\multicolumn{6}{c}{NGC\,1569}\\
\hline

Total &  $216\pm68$ & $772\pm244$ & $586\pm204$ & $3.69\pm1.74$ & $599\pm134$\\
Centre  & $44\pm14$ & $156\pm50$ & $172\pm60$ & $2.55\pm1.20$ & $599\pm134$\\
Outer Regions & $176\pm56$ & $306\pm97$ & $207\pm72$ & $8.54\pm4.03$ & $417\pm94$\\

\hline
\multicolumn{6}{c}{NGC\,3077}\\
\hline
Total &  $227\pm72$ & $260\pm83$ & $50\pm18$ & $45.56\pm21.67$ & $339\pm76$\\
Centre  & $1.3\pm0.4$ & $4\pm1$ & $72\pm30$ & $0.18\pm0.10$ & $547\pm131$\\
Outer Regions  & $114\pm36$ & $129\pm41$ & $41\pm15$ & $27.49\pm13.19$ & $338\pm76$\\

\hline
\multicolumn{6}{c}{NGC\,4214}\\
\hline

Total  &  $170\pm55$ & $182\pm59$ & $76\pm28$ & $22.40\pm11.05$ & $328\pm75$\\
Centre &  $13\pm4$ & $18\pm6$ & $37\pm18$ & $3.55\pm2.06$ & $373\pm91$\\
Outer Regions    & $32\pm10$ & $42\pm14$ & $162\pm66$ & $1.97\pm1.03$ & $365\pm85$\\

\hline
\multicolumn{6}{c}{NGC\,4449}\\
\hline

Total  &  $2214\pm700$ & $3255\pm1030$ & $228\pm79$ & $97.16\pm45.65$ & $384\pm86$\\
Centre &  $119\pm38$ & $183\pm58$ & $62\pm22$ & $19.05\pm9.02$ & $393\pm88$\\
Outer Regions   & $703\pm223$ & $1155\pm366$ & $245\pm86$ & $28.66\pm13.50$ & $406\pm91$\\

\hline
\multicolumn{6}{c}{NGC\,5253}\\
\hline

Total   & $318\pm101$ & $551\pm175$ & $374\pm131$ & $8.50\pm4.02$ & $417\pm94$\\
Centre  & $61\pm19$ & $115\pm36$ & $93\pm33$ & $6.57\pm3.12$ & $433\pm97$\\
Outer Regions    & $138\pm45$ & $219\pm73$ & $337\pm135$ & $4.11\pm2.11$ & $399\pm92$\\

\hline
\multicolumn{6}{c}{He\,2-10}\\
\hline

Total  &  $1784\pm578$ & $2482\pm813$ & $39\pm15$ & $463\pm234$ & $374\pm86$\\
Centre &  $21\pm7$ & $79\pm26$ & $27\pm10$ & $7.81\pm3.76$ & $614\pm139$\\
Outer Regions  & $373\pm119$ & $591\pm189$ & $47\pm17$ & $79.22\pm37.96$ & $399\pm90$\\

\hline
\end{tabular}

\end{minipage}
\end{table*}

\subsection{Non-detections}
\label{sec:nondiffuse}

\begin{figure}
\centering
\includegraphics[width=8cm]{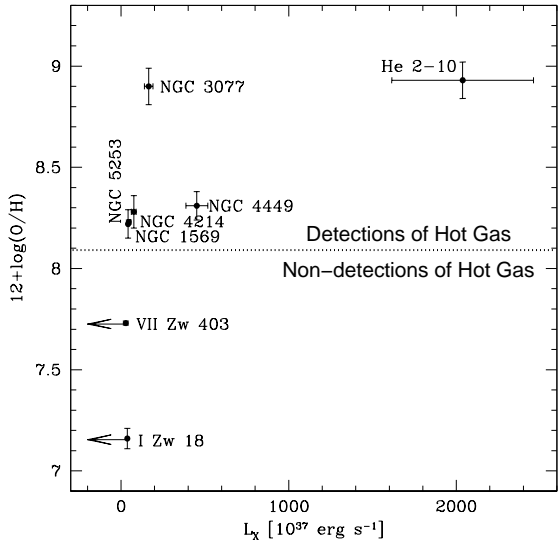}
\caption
{Oxygen abundance of \hii\ regions plotted against the X-ray
  luminosity of the diffuse component. Below a threshold of
  $12+\log(O/H)\sim8.1$ no hot gas is detected within the
  detection limits.}
\label{fig:lx_met}
\end{figure}

No diffuse X-ray emission is detected in I\,Zw\,18 and VII\,Zw\,403.
The integration times of both objects are the lowest of all the
galaxies in the sample and the detection limits of the unabsorbed
luminosities are of order $\sim 2-4\times 10^{38}$\,\lum\ (see
Paper\,I). This is about one order of magnitude lower than what is
measured for the faintest galaxy with detection
(Table\,\ref{tab:detect_gasb}). The 5--10 times smaller size of
I\,Zw\,18 and VII\,Zw\,403 (see Table\,\ref{tab:sample_props_opt}),
however, reduces the volume of these galaxies up to three orders of
magnitude. For this reason, the hot gas might be below the detection
threshold if it has similar properties to that found in the rest of
the sample. A second factor may be the metallicity of the gas, which
controls the X-ray emissivity \citep[see, e.g., ][]{sut93}. As shown
in Paper\,I, the metallicities of the hot gas are in agreement with
the metallicities derived directly from \hii\ regions. As I\,Zw\,18 and
VII\,Zw\,403 are the objects with the lowest metallicities in the
sample, this might push their X-ray luminosity below the detection
threshold. In particular, all galaxies where diffuse X-ray emission is
detected exhibit oxygen abundances in excess of $12+\log(O/H)\ga 8.1$
(see Fig.\,\ref{fig:lx_met}).

Alternatively, a higher absorbing column density or a very high
temperature can result in a non-detection with hot gas still being
present in large quantities (note that the fitted column densities for
the other objects are 2--4 times higher as compared to $1.5\times
10^{21}$\,\cden\ which are adopted for the determination of the
detection threshold). We also cannot exclude that hot gas in I\,Zw\,18
and VII\,Zw\,403 has not formed or that the hot gas might have blown
out of the gravitational potential of VII\,Zw\,403 (see
Sect.\,\ref{sec:out}). Alternatively, it might have cooled down
sufficiently to escape detection. The ages of the expanding \ha\
shells are indeed higher in I\,Zw\,18 and VII\,Zw\,403 (10--17\,Myr)
as compared to the other galaxies (1--10\,Myr), see
\citep{mar98,sil02}. Another possibility would be that the conditions
for developing shock fronts are different
\citep[see the models in][]{str02}. Deeper X-ray observations are
needed for a final assessment.

\subsection{Luminosities}

Hot gas is detected in NGC\,1569, NGC\,3077, NGC\,4214, NGC\,4449,
NGC\,5253, and He\,2-10. Their X-ray luminosities cover about two
orders of magnitude. The galaxies with the lowest diffuse X-ray
luminosities are NGC\,1569 ($L_{\rm X}=4.2\times 10^{38}$\,\lum) and
NGC\,5253 ($4.7\times 10^{38}$\,\lum, Table\,\ref{tab:detect_gasb}).
As shown in Sect.\,\ref{sec:morph}, these are the galaxies, where the
X-ray emission extends spatially beyond an \hi\ column density of
$10^{20}$\,\cden. As most of the X-rays are produced by shocked and
thermalised material it may be that there is no material in the halo
for the stellar ejecta to be hit, which reduces the X-ray luminosity.
In addition, those objects are the two galaxies with diffuse X-ray
emission showing the lowest metallicities ($\sim 20$ per cent solar,
see Table\,\ref{tab:sample_derived}). As the X-ray emissivity of a
given amount of material is a nearly linear function of metallicity
\citep{sut93} this could also be the cause for the low luminosity 
in these two objects. The galaxies where the diffuse X-ray emission
seems to be still confined to the extended \hi\ envelope show
luminosities of $7.6\times 10^{38}$\,\lum\ (NGC\,4214) and $4.5\times
10^{39}$\,\lum\ (NGC\,4449). These galaxies have a slightly larger
metallicity of $\sim 25$ per cent solar. In He\,2-10 the X-ray
emission extends about as far as the size of the \hi\ at
$10^{20}$\,\cden. It is the galaxy with the highest X-ray luminosity
($2.0\times 10^{40}$\,\lum) and metallicity (about solar). The only
galaxy with an X-ray luminosity that does not fit into the trend with
metallicity, shown in Fig.\,\ref{fig:lx_met}, is NGC\,3077 with a
metallicity of solar and a relatively low X-ray luminosity of
$1.7\times10^{39}$\,\lum.

\subsection{Temperatures}
\label{sec:temperature}

The temperature of the hot gas averaged over the total area is with
$\sim 2.0-3.5\times 10^{6}$\,K about the same for all galaxies, except for
NGC\,1569 ($T\simeq 7.2\times 10^{6}$\,K). As described in
\citet{hec93,hec02,str02}, temperatures $T$ in this range are 
due to shocks and are a direct function of shock velocities $v_{s}$:
$v_{s}=[(16/3)(kT/\mu)]^{0.5}$ ($k$: Boltzmann's constant, $\mu$: mean
mass per particle), resulting in $v_{s}\simeq 470$\,\kms\ for
$T=3\times 10^{6}$\,K. This value is similar to the derived mean
particle velocity of the hot, thermalised gas
(Table\,\ref{tab:common_gasparams}). Except for NGC\,4214, NGC\,4449,
and NGC\,5253, the temperature of the hot gas stored in the outer
regions of the galaxies is $\sim 2-3$ times lower than in their
centres (Table\,\ref{tab:detect_gasb}). As most of the X-rays emerge
from shocked gas this might be indicative of decelerating wind
velocities, enhanced mass-loading of cooler gas, or adiabatic cooling
in the halo of the dwarf galaxies under study.

\subsection{Absorbing column densities}
\label{sec:columndensities}

The absorbing column densities which were fitted to the individual
spectra are in the range of $\sim 3-7\times 10^{21}$\,\cden. This is
not in particularly good agreement with \hi\ data. Observations of the
21\,cm line of neutral hydrogen result in \hi\ column densities of
$0.1-2\times 10^{21}$\,\cden (see Fig.\,\ref{fig:HI_olays}). When the
absorbing column density is fixed to the \hi\ values, the fit to the
X-ray spectra becomes worse in the soft regime. Also the application
of non-equilibrium plasmas ({\sc XSPEC} model {\sc xsnei}), a
combination of two thermal plasma components, a thermal plasma with a
power law (to incorporate some faint emission from unresolved point
sources), or a variation of the $\alpha/Fe$ element abundance ratio
does not lower the absorbing columns substantially, neither does
fitting smaller regions (as done in Paper\,I) help.

A metallicity for the {\it absorbing} material which is higher than
that of the \hii\ regions can partly accommodate for this effect but
this is considered to be unlikely. An alternative explanation for the
discrepancy of measured versus fitted \hi\ column density might be
that the gas is more clumpy, or that the contribution by molecular
(and ionised gas) has been underestimated.

\subsection{Volume densities}
\label{sec:densities}

\begin{figure*}
\centering

\includegraphics[width=\textwidth]{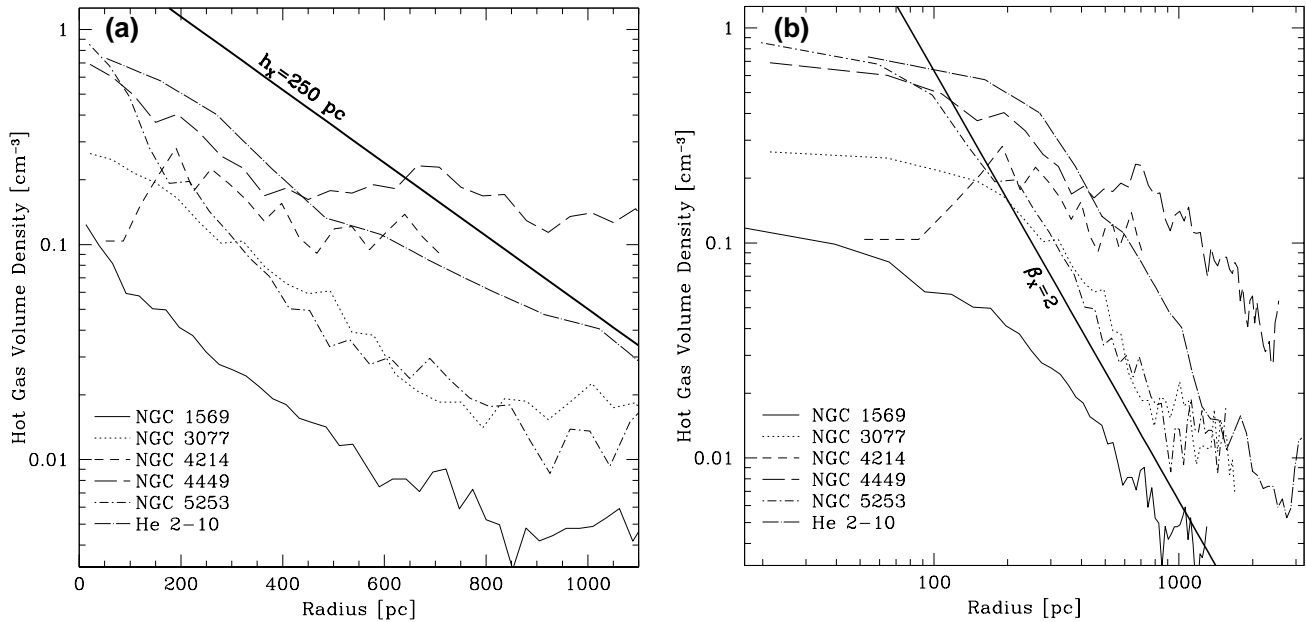}
\caption
{Azimuthally averaged volume density profile of the hot, X-ray
emitting gas in logarithmic {\bf (a)} and double logarithmic {\bf (b)}
units. For comparison, an exponential function with a scale length of
$h_{X}=250$\,pc and a power law with an index of $\beta_{X}=2$ are
displayed.}
\label{fig:common_dens}
\end{figure*}

From the fits to the diffuse X-ray emission listed in
Table\,\ref{tab:detect_gasb} we derive that the mean volume densities
of the hot gas are in the range of $n_{\rm e}\simeq 0.02-0.06$\,\vden\
averaged over each galaxy (Table\,\ref{tab:common_gasparams}). In
general, volume densities at their centres are higher by a factor of
$\sim 3$ as compared to the outskirts. With the help of the X-ray
surface brightness profiles (Fig.\,\ref{fig:profileboth}) and the
method described in OMW03 we constructed {\it volume density} profiles
assuming spherical geometries. Those profiles are shown in
Fig.\,\ref{fig:common_dens}. In general, exponential fits are
appropriate for the inner $\sim 400$\,pc whereas power law models
are better fits to the data at larger radii. Results of the fits are
listed in Table\,\ref{tab:xprofile_dens}. Except for NGC\,4449 the
exponential scale lengths of the volume densities are up to 50 per cent
larger as compared to the azimuthally averaged X-ray surface
brightness scale lengths (Table\,\ref{tab:profiles_fit}) and are in the
range of $150-320$\,pc. The power law index $\beta_{\rm X}$
(definition: $n_{e}\propto r^{-\beta_{\rm X}}$) for the outer parts of
the profiles are all within a range of $\beta_{\rm X}\sim1.5-2.0$ with
the only exception being NGC\,4214 ($\beta_{\rm X}=0.81$, see
Fig.\,\ref{fig:common_dens}). For comparison: a spherically-symmetric
(and also a conical), freely flowing wind has a power law index of 2
\citep{che85} which is very close to the measured values 
(see also the discussion in Sect.\,\ref{sec:out}).

\begin{table}
\centering
\caption{Scale lengths ($h_{\rm X}$ in pc) for exponential 
functions to the inner radii ($r\la 350-600$\,pc) of the volume
density profiles and spectral indices for power law fits
($\beta_{\rm X}$) at larger radii. }
\begin{tabular}{@{}lcc@{}}

\hline
\multicolumn{1}{l}{Galaxy}&\multicolumn{1}{c}{$h_{\rm X}$}&\multicolumn{1}{c}{$\beta_{\rm X}$}\\
\hline

NGC\,1569 &  $233\pm  \phn7$ & $1.72\pm 0.19$\\
NGC\,3077 & $274\pm 10$      & $1.85\pm 0.14$\\
NGC\,4214 & $316\pm 67$      & $0.81\pm 0.24$ \\
NGC\,4449 & $272\pm 17$      & $1.46\pm 0.07$ \\
NGC\,5253 & $142\pm \phn7$   & $1.85\pm 0.09$ \\
He\,2-10 & $279\pm 38$      & $1.96\pm 0.09$ \\

\hline
\end{tabular}
\label{tab:xprofile_dens}
\end{table}

\subsection{Pressures}
\label{sec:pressures}

The pressures of the hot gas are in the range of $P/k\simeq 1-3\times
10^{5}$\,K\,\vden\ for all galaxies. The ISM in our Milky Way exhibits
pressures of at most a few times $10^{4}$\,K\,\vden\ \citep[see, e.g.,
][]{wol03}. If this value was also typical for dwarf galaxies
undergoing a starburst, then the hot gas has an overpressure and will
expand. This naturally explains the expanding \ha\ shells in the
sample of galaxies under study \citep[see,
e.g.,][]{mar98,mar95,mar96}. The pressure of the hot gas has a
gradient from the centre of the galaxies toward the outer regions (see
Table\,\ref{tab:common_gasparams}); the pressures in the centres are
on average $\sim 5$ times higher than those in the outer regions. This
is due to the combined effect of both, lower volume densities and
lower temperatures in the outskirts.

\subsection{Masses and mass-loading of the hot gas} 
\label{sec:massload}

The masses of the hot gas ($M_{\rm X}$) within the galaxies range from
$\sim 2$ to $20\times10^{6}$\,\msun. Most of the mass is stored in the
outer regions of the galaxies, which is a consequence of the much
larger volume as compared to the centres. The mass of the hot gas is
about 1 per cent of the \hi\ content of the galaxies (see
Tables\,\ref{tab:common_gasparams} and \ref{tab:sample_derived}).

We will now compare the mass injection from massive stars ($M_{\rm
ej}$) to $M_{\rm X}$. For this purpose, we use the stellar evolution
synthesis models published by \citet{lei99} ({\sc
STARBURST99}\footnote{\sl
http://www.stsci.edu/science/starburst99/}). In their figure\,10, they
show the mass-loss of a stellar population undergoing a continuous
starburst at a rate of 1\,\msun\,yr$^{-1}$. Throughout this discussion
we will use the models based on a Salpeter IMF with an upper mass
cutoff of 100\,\msun\ and a lower mass cutoff of 1\,\msun. As
described in \citet{mar98} and \citet{men99}, expanding shells in most
of the starburst galaxies in our sample have a kinematic age of $\sim
10^{7}$\,yr. This expansion was mainly driven by the kinetic energy
input of SNe, which explode $\sim 4\times 10^{7}$ yr after the
progenitor stars were created. For this reason, we evaluate the
integrated mass loss at an age of $5\times 10^{7}$\,yr. At this age
the models are virtually independent of metallicity. To derive the
final mass loss of the stellar population in the galaxies of our
sample, we scale the {\sc STARBURST99} models at that point of stellar
evolution by the current SFRs listed in
Table\,\ref{tab:sample_derived} ($SFR_{H\alpha}$: based on \ha\ and
$SFR_{\rm FIR}$: based on FIR measurements). The resulting $M_{\rm
ej}$ is then used to compute the corresponding mass-loading factor
$\chi_{\rm ml}$ which we define as:

\begin{equation}
\chi_{\rm ml}=\frac{M_{\rm X}}{M_{\rm ej}}-1
\end{equation}

$\chi_{\rm ml}$ therefore corresponds to the amount of entrained
material relative to the mass of the stellar ejecta. The mass-loading
factors $\chi_{\rm ml}$ for our sample of galaxies are estimated to
be: $-0.6/0.9$ (NGC\,1569), $3.0/1.0$ (NGC\,3077), $0.7/1.7$
(NGC\,4214), $5.2/5.4$ (NGC\,4449), $1.3/0.9$ (NGC\,5253), and
$4.3/0.8$ (He\,2--10) where the first number is derived from
$SFR_{H\alpha}$ and the second figure using $SFR_{\rm FIR}$. These
estimates show that at least about the same amount as the stellar
ejecta itself was entrained by the superwind. This material must have
been shocked and/or boiled off the surrounding, cooler ISM
\citep[cf. the models of][]{wea77,str02}. In the process, the hot gas 
was cooled down to the measured temperatures of a few times
$10^{6}$\,K (see Table\,\ref{tab:detect_gasb}), which boosts the X-ray
emissivity of the hot gas \citep[see the cooling functions described
in][]{sut93}. We note, however, that all of the above values were
derived for a volume filling factor of unity and for a metallicity for
each galaxy based on that of its \hii\ regions. A lower filling
factor, as suggested by the calculations of \citet{str00} or
supersolar metallicities
\citep[e.g., as derived for NGC\,1569 by][]{mar02} will reduce the
need for mass loading.

\subsection{Thermal energies}

The thermal energy ($E_{th}$) of the hot gas is a lower limit to the
mechanical energy input of strong stellar winds and SNe. Only in the
absence of cooling and under the condition of full thermalisation both
energies should be equal. $E_{th}$ depends linearly on the mass of the
hot gas and on its temperature. The temperatures of the galaxies in
the sample are very similar (see Sect.\,\ref{sec:temperature}). For
this reason, the thermal energies vary, as do the masses, over an
order of magnitude: $E_{th}\simeq 2-30\times 10^{54}$\,erg. Most of
the thermal energy is stored in the outer regions rather than in the
centres of the galaxies which is due to the much larger volume at
larger radii. The total thermal energy corresponds to the kinetic
energy output of $\sim 2000-30000$ SNe. This figure is confirmed by
\ha\ spectroscopy of expanding shells (see Sect.\,\ref{sec:out}). 
According to the STARBURST99 models, a stellar cluster with a combined
mass of $10^{6}$\,M$_{\sun}$ releases an energy of $\sim 10^{55}$\,erg
(in the form of SN explosions and stellar winds) which is similar to
what we derive for the thermal energies of the hot thermal gas. This
is in agreement with the detection of super-stellar clusters in the
starburst galaxies (see Paper\,I for references).

\subsection{Cooling times, mass deposition rates and velocities}

We derive the shortest cooling times ($\sim50$\,Myr) for NGC\,3077 and
He\,2-10. The cooling time is much longer for NGC\,1569 and NGC\,5253
(590 and 370\,Myr respectively) and NGC\,4214 and NGC\,4449 fall in
between (80 and 230\,Myr). In general, the cooling times in the
centres of the galaxies are shorter than in their outskirts
(Table\,\ref{tab:common_gasparams}). The faster cooling toward the
centres of the galaxies, however, does not result in a larger mass
deposition rate $\dot{M}_{\rm cool}$ (the rate at which the amount of
coronal gas cools down). Indeed, $\dot{M}_{\rm cool}$ is mostly lower
toward the centres than toward the outer regions. This can be
understood by the fact that most of the hot gas is stored in the
outskirts rather than in the centres of the galaxies.

\begin{figure}
\centering
\includegraphics[width=8cm]{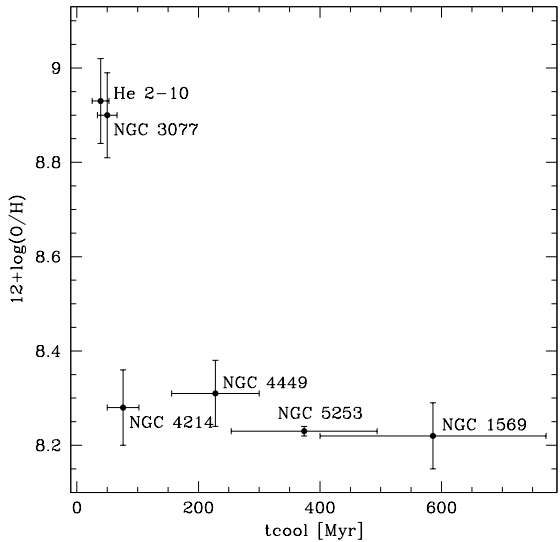}
\caption
{The \hii\ metallicities of the galaxies plotted against the cooling
time of the hot gas.}
\label{fig:correl_tcool}
\end{figure}

The ability of the hot gas to cool down is an important parameter for
the evolution of the galaxies. Dwarf galaxies, in general, have a
lower metallicity as compared to more massive spirals. The connections
between the metallicities and the cooling times are shown in
Fig.\,\ref{fig:correl_tcool}. In spite of the small difference in
metallicity between most of the galaxies in the sample, a trend can be
observed in the sense that the cooling is faster at larger
metallicities.

The mean particle velocities fall between $\sim 300$ and
$600$\,\kms. Differences between velocities for the central and outer
regions correlate with their temperatures. This again
indicates that at a given temperature the shock velocities are about
the same for all dwarf galaxies considered here (see
Sect.\,\ref{sec:temperature}).


\section{Correlation of observational parameters}
\label{sec:correl}

\begin{figure*}
\centering
\includegraphics[width=18cm]{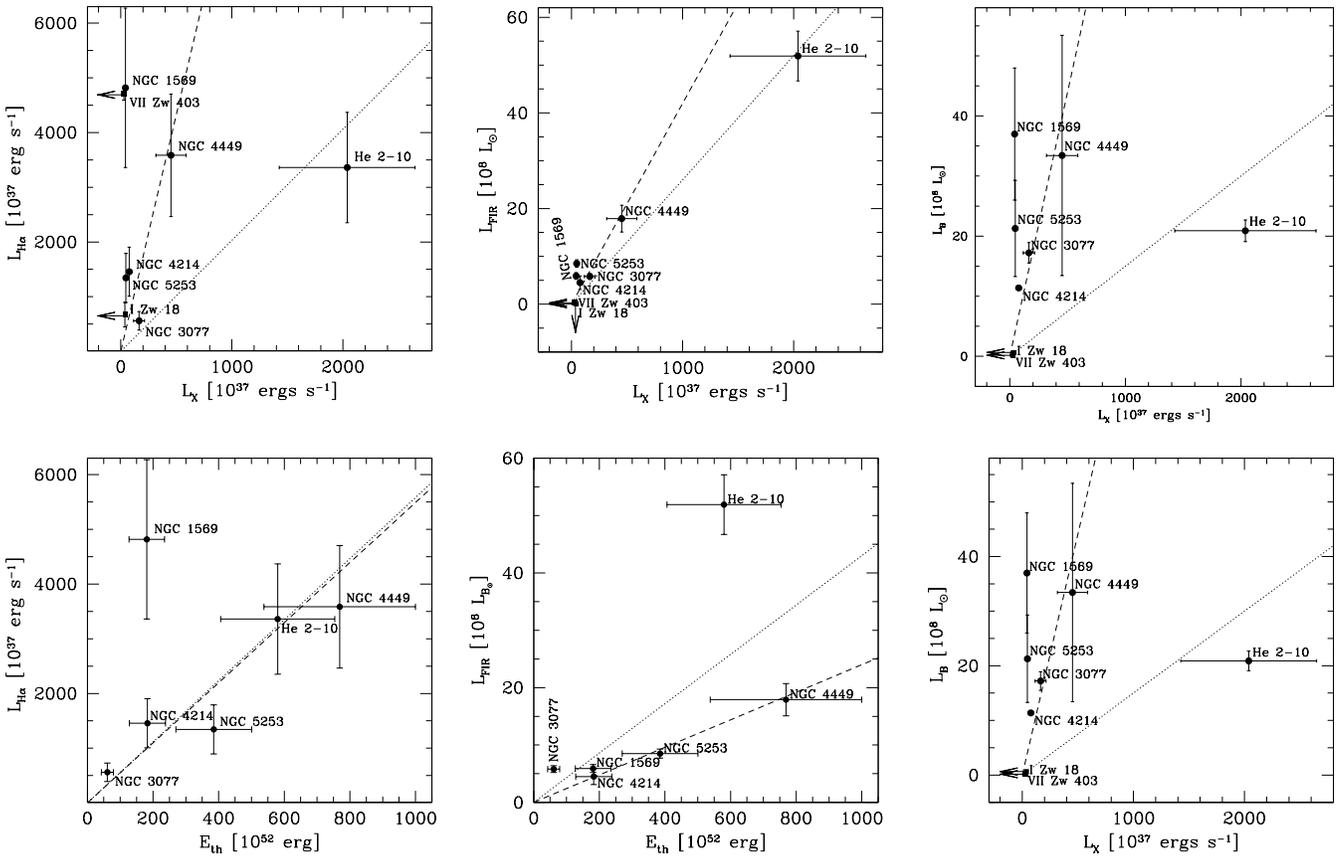}
\caption
{The SFR tracers \ha, FIR emission, and blue luminosity as a function
of the X-ray luminosity ({\bf upper panels}) and thermal energy ({\bf
lower panels}) of the hot gas. Least squares fits to the graphs are
plotted as {\bf dashed} ({\bf dotted}) lines including (excluding)
He\,2-10. The slopes of these fits are listed in
Table\,\ref{tab:params_fit_diffuse}.}
\label{fig:correl_starformation}
\end{figure*}

Starburst galaxies are characterised by higher SF rates (SFRs) as
compared to quiescent galaxies. The combined effect of type II SNe and
strong stellar winds then lead to the heating of the ambient medium to
coronal temperatures. Consequently, SF tracers such as the \ha\
luminosity (from \hii\ regions), the FIR luminosity or the blue
luminosity are expected to be related to hot gas parameters. Whereas
the \ha\ and FIR luminosities are signatures of current SF, the blue
luminosity represents young stars which can live up to a few hundred Myr
\citep[e.g., ][]{lei99}. It is therefore a tracer of recent rather
than current SF. In Fig.\,\ref{fig:correl_starformation} we show
correlations of these SF tracers with the X-ray luminosity $L_{\rm X}$
of the hot gas and its thermal energy $E_{\rm th}$ (all fits are
forced to incorporate the origin). Best results are obtained for fits
excluding He\,2-10 and, after converting the \ha\ and FIR luminosities
into SF rates (see Sect.\,\ref{sec:sample_props_other}), the
correlations are listed in Table\,\ref{tab:params_fit_diffuse} (note
that the absorbing column density of He\,2-10 listed in
Table\,\ref{tab:common_gasparams} is larger for the entire galaxy than
for its centre and the correction for absorption might therefore
overestimate its X-ray luminosity by an order of magnitude; see also
the discussion in Sect.\,\ref{sec:point_sfr}). In particular, the
trends with both, the
\ha\ and FIR SFRs are the same within the errors. Galaxies without
diffuse X-ray emission have lower blue luminosities \citep[see
also][]{ste98}, metallicities (Fig.\,\ref{fig:lx_met}), and SFRs than
those with hot gas (but see the relatively large \ha\ luminosity of
VII\,Zw\,403). 

Note that the scatter of the \ha-$L_{\rm X}$ plot
(Fig.\,\ref{fig:correl_starformation}[left]) seems to be larger than
what is suggested by the surface brightness profiles
(Sect.\,\ref{sec:morph} and Fig.\,\ref{fig:profileboth}). Both, the
\ha\ and X-ray surface brightness profiles are very similar 
which can be explained by the properties of the developing shock
fronts. In Fig.\,\ref{fig:correl_starformation}(left), however, the
total \ha\ emission is shown, i.e., the \hii\ regions and the more
diffuse \ha\ emission. The former has been excluded from the surface
brightness analysis.

\begin{table*}

\caption
{Star formation tracers as a function of the X-ray luminosity $L_{\rm
X}$ and the thermal energy $E_{\rm th}$ of the diffuse, hot gas. Shown
are the fits for the sample (see
Fig.\,\ref{fig:correl_starformation}). The quantities in the header
equal the proportionality constants in the table multiplied by the
quantities in the first columns. Values in square brackets exclude
He\,2-10 from the fits ($SFR_{\rm H\alpha}$ and $ SFR_{\rm FIR}$ are
the star formation rates derived from \ha\ and FIR luminosities,
respectively, $L_{\rm B}$ is the blue luminosity).}
 
  \begin{tabular}{@{}lccc@{}}
\hline
                  & $SFR_{\rm H\alpha}$             & $ SFR_{\rm FIR}$ & $L_{\rm B}$ \\
                  & [$10^{-4}$\,\msun\,yr$^{-1}$]   &[$10^{-4}$\msun\,yr$^{-1}$]& [$10^{5}\,L_{B, \sun}$]\\
\hline
$L_{\rm X}    $   &  $1.81\pm1.10$    & $4.48\pm3.45$    & $15\pm11$\\

[$10^{37}$\,\lum] &  $[7.72\pm4.29]$  & $[7.24\pm1.38]$  & $[88\pm39]$\\
$E_{\rm th}    $  &  $1.15\pm0.17$    & $1.94\pm0.64$    & $10.4\pm2.9$\\

[$10^{52}$\,erg]  &  $[1.32\pm0.54]$  & $[1.03\pm0.19]$  & $[13.4\pm5.2]$\\
\hline
  \end{tabular}
    \label{tab:params_fit_diffuse}
\end{table*}

\begin{figure*}
\centering
\includegraphics[width=18cm]{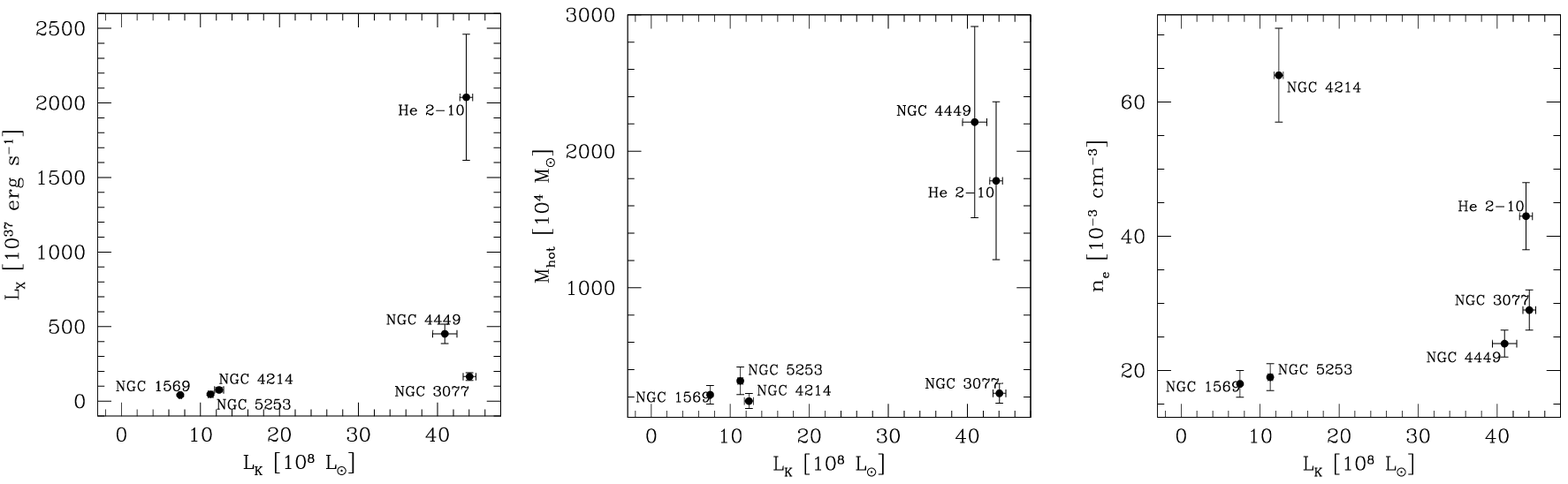}
\caption
{X-ray luminosity, mass, and density of the hot gas plotted
  against the $K$-band luminosity of the host galaxy.}
\label{fig:correl_lk}
\end{figure*}

\begin{figure*}
\centering
\includegraphics[width=18cm]{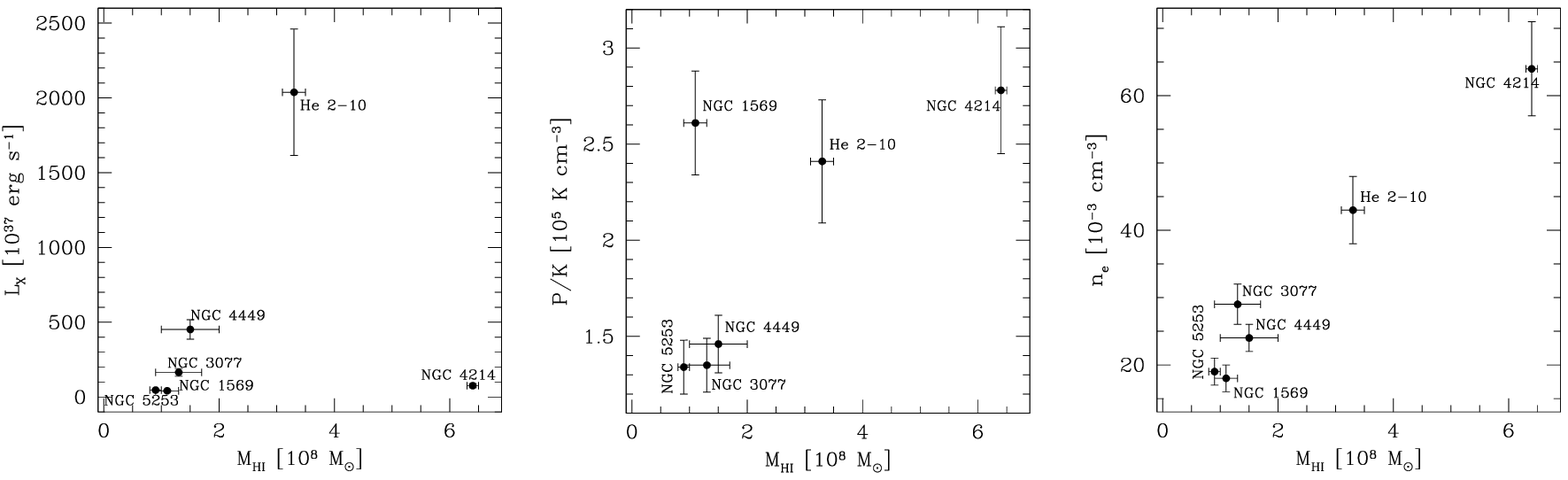}
\caption
{X-ray luminosity, pressure, and density of the hot gas plotted
  against the \hi\ mass of their respective hosts.}
\label{fig:correl_hi}
\end{figure*}

The $K$-band luminosities of the galaxies
(Table\,\ref{tab:sample_derived}) may be used as a tracer for their
total mass as it is dominated by low-mass and/or old stellar
populations. In Fig.\,\ref{fig:correl_lk} we plot the X-ray
luminosity, the thermal energy, and the density of the hot gas as a
function of this parameter. Some tentative trends are visible in the
sense that the X-ray luminosities and the thermal energies are larger
for more massive objects. Also the density of the hot gas appears to
be somewhat larger with increased $K$-band luminosities. The
parameters of the hot gas certainly depend on their ambient, gaseous
medium. To check this, we plot the X-ray luminosities, the densities,
and the pressures of the hot gas as a function of the \hi\ masses in
Fig.\,\ref{fig:correl_hi}. Here, the density again appears to be a
function of the \hi\ mass, i.e., a large \hi\ mass is correlated with
a higher density of the hot gas. As the temperatures of the hot gas
are relatively uniform over the sample, this trend is also visible
when plotting the pressures versus \hi\ masses. The X-ray luminosity,
however, seems not to be related to this parameter.

\section{Outflow}
\label{sec:out}

We will now discuss the question whether the energetic input of SF in
the galaxies may have resulted in gas removal or not. Theoretical
models in the literature try to describe the conditions for this
scenario
\citep[e.g.,][]{mac88,mac99,fer00,sil01}. The escape velocity of the
galaxy is a crucial parameter and has to be exceeded by any superwind
in order to remove gas from the gravitational well. The escape
velocity depends on the mass of the system. It is very difficult,
however, to make an accurate estimate of the total dark matter
content. Especially for our sample, where at least three galaxies are
interacting with one or more companion galaxies (I\,Zw\,18, NGC\,3077,
NGC\,4449) it is virtually impossible to derive a reliable rotation
curve from \hi\ measurements and therefore to correctly estimate the
shape and depth of the gravitational potential. These galaxies are
characterised by close, large tidal tails visible in \hi\ (see
Sect.\,\ref{sec:morph}). In addition, the loss of ISM can only take
place if the hot gas moves supersonically when reaching the
scale length of the cooler, surrounding ISM
\citep{sil01}. If so, the hot gas will be re-accelerated and cannot be
retained by its host. If this is not the case, the gas will cool down
in the galaxy's halo and eventually rain back on its disc.

\begin{figure*}
\centering
\includegraphics[width=18cm]{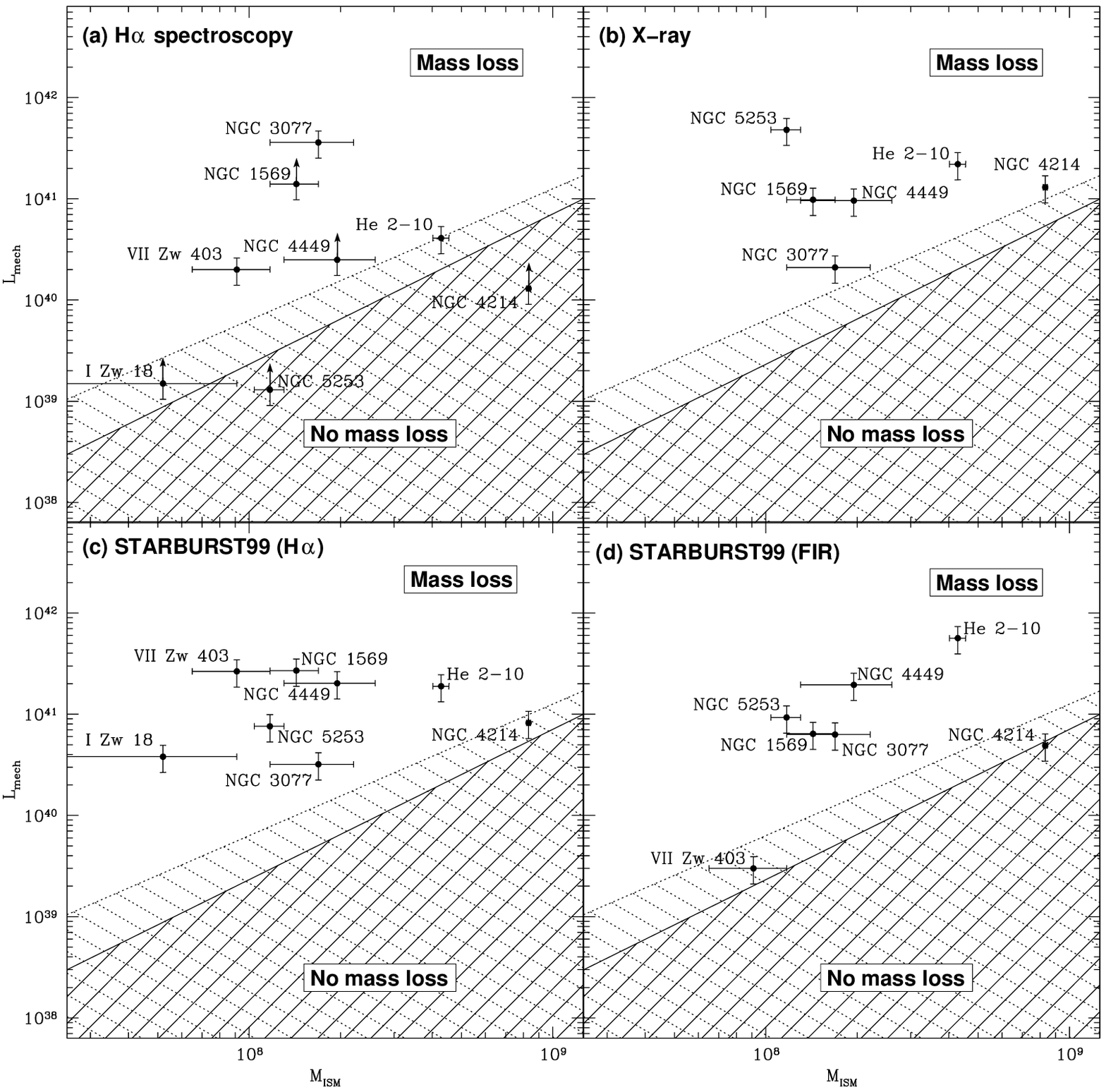}
\caption
{Mechanical luminosities of the starbursts within the galaxies as a
function of their ISM mass (He corrected \hi\ masses). Shown are
theoretical thresholds taken from \citet{sil01} for non-rotating
objects; for objects falling within the region outside of the {\it
shaded areas}, material is ultimately lost to the IGM, starburst
galaxies within the shaded region retain their ISM. The {\it solid}
threshold was computed for an IGM pressure of $P_{\rm
IGM}/k=100$\,\vden\,K and the {\it dotted} threshold for $P_{\rm
IGM}/k=1$\,\vden\,K. The thresholds are lower when the galaxies are
rotating. Mechanical luminosities are derived from {\bf (a)} \ha\ long
slit spectroscopy (lower limits are computed for a lower ambient gas
density limit of 0.1\vden), {\bf (b)} X-ray parameters, {\bf (c)} {\sc
STARBURST99} models scaled by $SFR_{H\alpha}$, and {\bf (d)} {\sc
STARBURST99} models scaled by $SFR_{FIR}$.}
\label{fig:silich}
\end{figure*}

For the highly disturbed galaxies, one would need to elaborate
detailed hydrodynamical simulations for a correct description of the
fate of the superwinds. This, however, is beyond the scope of this
paper. Instead, we discuss the probabilities for escape in the light
of the model described by \citet{sil01} \citep[with galaxy dynamics
based on][]{koo92,mac99}. They derive the threshold for the energy
input of the starburst required to overcome the gravitational
potential based on the ISM mass and the pressure of the surrounding
IGM (with a scaling law for dark matter of: $M_{DM}=3.47\times
10^{8}(M_{\rm ISM}/10^{7} M_{\sun})^{0.71} M_{\sun}$). If those
thresholds are exceeded, the gas is ultimately lost to the IGM. Their
results are shown in Fig.\,\ref{fig:silich}. Using their thresholds,
the graph is split into an area where the galaxies do not suffer mass
loss and an area where mass can escape to the IGM. The plotted
thresholds correspond to a galaxy without rotation and is therefore an
upper limit for the injected mechanical luminosity to guarantee
retention of the gas; when rotation is introduced to their models, the
thresholds are lowered. The dotted line in Fig.\,\ref{fig:silich} was
computed for an IGM pressure of $P_{IGM}/k=100$\,K\,cm$^{-3}$ and the
solid line for $P_{IGM}/k=1$\,K\,cm$^{-3}$. We plotted the galaxies in
our sample in this figure. The mechanical luminosities in
Fig.\,\ref{fig:silich}(a) are based on \ha\ long slit spectroscopy of
\citet{mar98} (I\,Zw\,18, NGC\,1569, NGC\,3077, NGC\,4214, NGC\,4449,
NGC\,5253), \citet{men99} (He\,2-10), and \citet{sil02}
(VII\,Zw\,403). The mechanical luminosities derived by \citet{mar98}
depend linearly on the ambient density of the gas. In OMW03, densities
were derived for individual superbubbles in the case of NGC\,3077 and
the total mechanical luminosity shown here is corrected
accordingly. For the rest of the sample we use a mean density of
0.1\,cm$^{-3}$. Note however, that the densities of the ambient gas
can be an order of magnitude higher (as in the case NGC\,3077, see
OMW03), and we plot the corresponding luminosities as lower limits in
Fig.\,\ref{fig:silich}(a). The ISM mass for all galaxies is based on
the \hi\ masses taken from Table\,\ref{tab:sample_derived} and are
corrected for the contribution of helium by a multiplication of
1.3. As it is difficult to correct for the amount of molecular gas, we
do not account for it; if the molecular medium would be as massive as
the atomic phase (an extreme scenario) the location of the points on
the logarithmic scale used in Fig.\,\ref{fig:silich} would shift by
0.3 to the right.

In addition to the mechanical luminosities derived by \ha\
spectroscopy, we estimate mechanical luminosities from our X-ray
observations in a different, independent manner. The timescale for
mass injection was taken as the scale length of the hot gas divided by
the mean velocity per particle (see Tables\,\ref{tab:profiles_fit} and
\ref{tab:common_gasparams}). Subsequently, the thermal energies were
divided by the resulting timescales. The luminosities thus obtained
are plotted as a function of ISM mass in Fig.\,\ref{fig:silich}(b).

Finally, we estimate the mechanical luminosities liberated by the
starbursts on the basis of the {\sc STARBURST99} models \citep[][their
fig.\,112, see Sect.\,\ref{sec:massload} for the parameters of the
starburst which we adopt here]{lei99}. Scaling these models by the
SFRs given in Table\,\ref{tab:sample_derived} and using the asymptotic
solution of the {\sc STARBURST99} model (note that the mechanical
luminosity stays approximately constant for starburst ages larger than
the timescale needed for the first SN to explode), we derive the
mechanical luminosities displayed in panels (c) and (d) of
Fig.\,\ref{fig:silich}. The data in Fig.\,\ref{fig:silich}(c) is based
on the SFRs derived on the basis of \ha\ measurements
($SFR_{H\alpha}$) and the data in Fig.\,\ref{fig:silich}(d) are
derived for the SFRs converted from the FIR luminosities of the
galaxies ($SFR_{\rm FIR}$).

Judging from the mechanical luminosities derived on the basis of \ha\
spectroscopy (Fig.\,\ref{fig:silich}[a]), gas loss is predicted for
VII\,Zw\,403, NGC\,3077, NGC\,1569, and NGC\,4449. He\,2-10,
I\,Zw\,18, NGC\,4214, and NGC\,5253 are not necessarily losing
material to the halo given their energetics. Those galaxies can exceed
the theoretical thresholds, however, if they rotate fast enough, the
ambient densities are higher than the adopted 0.1\,\vden\ (increasing
the mechanical luminosities), or the pressure of the IGM is relatively
low. According to the mechanical luminosities estimated from the
diffuse X-ray emission (Fig.\,\ref{fig:silich}[b]), the starbursts in
all galaxies exhibit a sufficient energy deposition rate to push the
hot gas away from the galaxies independent of their rotational
velocities. The results based on the {\sc STARBURST99} models yield a
very similar picture, and only the superwinds in NGC\,4214 (and maybe
VII\,Zw\,403) might be too weak to force gas loss if their rotation is
not significant. Fig.\,\ref{fig:silich} therefore strongly suggests
that the energy provided by the starburst is sufficient for these
galaxies to develop a superwind, leading to a partial loss of the ISM.
The observed overpressures of the hot gas as compared to the
surroundings support this scenario (Sect.\,\ref{sec:pressures}).

As a caveat, however, the ISM must not be very extended for gas loss
to occur \citep[see, e.g, ][]{sil98,derc99,sil01}. As seen in
Fig.\,\ref{fig:HI_olays}, NGC\,3077 and NGC\,4449 are surrounded by
large tidal tails. The same is true for I\,Zw\,18
\citep[see the \hi\ map of][]{vze98}. Inspection of the \hi\ and
X-ray morphologies reveals that the diffuse X-ray emission of
NGC\,3077 is more extended toward the north than toward the south
following the largest \hi\ density gradient. Toward the south of
NGC\,3077, the \hi\ column density stays constant at a level of $\sim
1.5\times 10^{21}$\,cm$^{-2}$ (see OMW03). Once the hot gas reaches the
scale length of the \hi\ while maintaining supersonic speed, it will
ultimately be lost. As discussed in OMW03 this is likely to happen
toward the north of NGC\,3077. Gas streaming to the south, however,
will likely cool down before reaching the low density IGM beyond the
large \hi\ tidal tail (see also Fig.\,\ref{fig:HI_olays}) unless SF
continues at a high level on timescales exceeding the cooling time.

\begin{figure*}
\centering
\includegraphics[width=15cm]{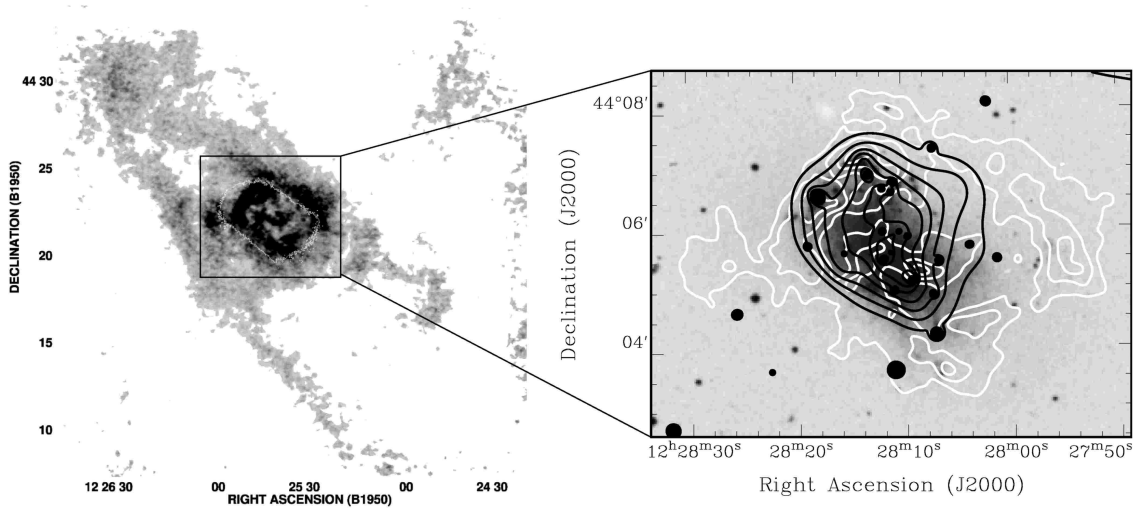}
\caption
{{\bf Left:} \hi\ map of NGC\,4449 and its surrounding tidal tails.
  The optical counterpart is outlines as {\bf white contours}. {\bf
    Right}: Optical DSS image of NGC\,4449. Overlaid are contours of
  the \hi\ (blow-up of left image: {\bf white}) and the X-ray
  emission ({\bf black}). The \hi\ data was kindly provided by
  \citet{hun99}.}
\label{fig:n4449_HI}
\end{figure*}

The case of NGC\,4449 is somewhat different. Its \hi\ is distributed
in a ring (mean \hi\ column density: $\sim 5\times 10^{20}$\,\cden)
tightly surrounding the optical counterpart with tidal tails connected
to the ring stretching outward (see Figs.\,\ref{fig:n4449_HI} and
\ref{fig:HI_olays}). Currently, the hot gas streams into the interior
of the ring which is the region of lowest \hi\ column
density. Depending on how much material can accumulate in this region
before the hot gas cools down, the superwind may break through this
'wall'. Even if the hot gas extends further out, it still has to
provide the work needed to push the gas aside which is stored in the
tidal tails. An escape along the line-of-sight is possibly not a
solution as even within the ring low density \hi\ emission is being
observed over a large velocity range (\citealt{hun99}; see also
\citealt{sum03}).

No current interaction or apparent tidal tails are observed for
NGC\,4214. Even though the hot gas is concentrated in the central
region (see Fig.\,\ref{fig:HI_olays}), it is very likely that the hot
gas escapes along the line-of-sight which corresponds more or less to
the z--axis of this object (note the relatively face-on appearance of
NGC\,4214).

In summary, dwarf galaxies have the potential to release metals to the
IGM in the form of powerful galactic superwinds. This is corroborated
by their large mechanical energy inputs in a relatively shallow
gravitational potential as well as by the overpressures of the hot
gas. The main trigger for starbursts, however, is galaxy-galaxy
interactions which produces extended tidal tails, and any low density
neutral envelope may provide a barrier for outflow to occur,
containing the gas for times larger than the cooling time scale and
thus preventing the complete loss of ISM to the IGM.


\section{Summary}
\label{sec:summary}

In Paper\,I we presented a detailed discussion of the individual X-ray
properties as observed with the {\it Chandra X-ray observatory} of a
sample of eight dwarf galaxies which are in a starburst phase,
I\,Zw\,18, VII\,Zw\,403, NGC\,1569, NGC\,4214, NGC\,4214, NGC\,4449,
NGC\,5253, and He\,2-10. Here we present an inter-comparison of their
X-ray properties as well as comparisons with observations at other
wavelengths. Our results can be summarised as follows:

\begin{itemize}

\item X-ray point sources (detected within the optical or \hi\ extent
  of an individual galaxy) are mainly observed close to the centres of
  the galaxies; only a few of them are found in their outer
  regions. In general, they are located close to bright \hii\ regions,
  the rims of expanding superbubbles, or UV bright young stellar
  clusters. The type of X-ray emission mechanism model (power law,
  thermal plasma, or black body) of the point sources appears to be
  unrelated to their location.

\item The number of point sources correlates with the star formation
  rates of the galaxies and, with less significance, with their blue
  luminosities. The cumulative luminosity function of the point
  sources exhibits a very shallow power law index of $0.24$. This
  value is about a factor of $2-4$ lower than that of more massive
  starbursts and about 6 times lower than that of non-starburst
  galaxies. Ten ultraluminous X-ray sources (X-ray luminosities
  $>10^{39}$\,\lum) are detected in three galaxies (I\,Zw\,18,
  NGC\,4449, He\,2-10). Each of these galaxies is part of a current
  interaction or merger.

\item Diffuse X-ray emission emerging from the coronal gas in
  superwinds is detected in six out of eight galaxies. The
  non-detection in I\,Zw\,18 and VII\,Zw\,403 can be understood as
  being a result of their smaller sizes and/or their extremely low
  metallicities as compared to the other galaxies in the sample. We
  cannot rule out that any hot gas that might have been there at
  earlier times has either vented into the halo or has cooled so as to
  become undetectable.

\item Superwinds traced by the diffuse, coronal gas develop along the
  steepest declining gradient in the ISM as observed in \hi. A higher
  temperature is observed in the centres of the galaxies as compared
  to that in their outskirts. For about half of the sample, the X-ray
  emission extends well beyond the \hi\ component whereas in the other
  half the hot gas appears to be still enclosed in an \hi\ envelope.
  On a global scale the azimuthally averaged \ha\ and X-ray surface
  brightness profiles are very similar and are well fitted by
  exponential functions with scale lengths $\sim 100-200$\,pc
  (NGC\,4449: $\sim 550$\,pc). This may be a consequence as that on
  global scales photoionisation and shocks of the \ha\ features as
  well as the thermalisation of the X-ray emitting gas are all due to
  the impact of massive stars on the ISM distributed over the discs of
  the starburst galaxies. On local scales the X-ray and \ha\ emission
  are either slightly offset or show the morphology of expanding \ha\
  shells filled with coronal gas.

\item Volume densities of the hot gas are in the range 
  $0.02-0.06$\,\vden, pressures vary within
  $1-3\times10^{5}$\,K\,cm$^{-3}$, thermal energies are $\sim
  2-30\times10^{54}$\,erg, and masses $2-20\times10^{6}$\,\msun. Thus,
  compared to their \hi\ content, about 1 per cent of the ISM is in
  the form of hot, coronal gas. The pressures involved are higher than
  those for the other components of the ISM. This is the most likely
  force driving the expansion. Mass-loading of a factor of $1-5$ is
  needed for the galaxies.

\item The X-ray luminosities and the cooling times of the hot gas
  appear to be related to the metallicities of their host galaxies.
  Low-metallicity objects have longer cooling times. Furthermore,
  galaxies with the largest mass-loading factors exhibit the shortest
  cooling times and the highest mass deposition rates.

\item Correlations are found for the \hi\ mass of the galaxies 
    with the pressures and densities of the hot gas. These parameters
    appear to increase with increasing \hi\ mass, the hydrodynamics of
    which appear to control the confinement of the hot gas. Such a
    correlation is only tentatively visible when plotting the hot gas
    parameters against the $K$-band luminosities (used as a tracer for
    total mass of the galaxies). In addition, the X-ray luminosities
    and thermal energies of the hot gas may be used to trace star
    formation in the galaxies. A larger sample, however, is needed to
    strengthen this correlation.

\item The mechanical energies, as derived from \ha\ and X-ray observations,
 are large enough to overcome the gravitational potential of their
 hosts. In combination with the fact that the current star formation
 is still high and that the hot gas shows an overpressure compared to
 the ambient ISM, we conclude that outflows are likely and that the
 hot, $\alpha$ elements enriched gas component is lost to the
 IGM. When large scale, low density \hi\ tidal tails are present,
 however, the hot gas appears to be confined within those tidal
 structures.

\end{itemize}

Dwarf galaxies that are undergoing a starburst are important objects
to study due to their potential contribution to the metal enrichment
of the IGM at large look-back times. The number of dwarf galaxies in
the Universe may be large enough to move significant amounts of matter
(and thus metals) to the intergalactic medium. Here we have shown that
outflows are possible in a number of nearby dwarf starbursts; however
some fraction of galaxies will retain their hot gas due to the
hydrodynamical drag of extended envelopes of cooler material. Deeper
observations of a larger sample of dwarf (starburst) galaxies will be
necessary to determine if and in which proportion these objects
contribute to the metal enrichment of the Universe.

\section*{Acknowledgements}
We would like to thank Dominik Bomans, John Cannon, Deidre Hunter,
Chip Kobulnicky, Vince McIntyre, and Stefanie M{\"u}hle for providing
optical, \ha\, and \hi\ images of the galaxies. In particular we thank
Crystal Martin for providing some optical images as well as for
valuable discussions on NGC\,3077 and He\,2-10. We are also grateful
to David Strickland for the critical reading of the manuscript. JO
acknowledges the Graduiertenkolleg 118 'The Magellanic System, Galaxy
Interaction, and the Evolution of Dwarf Galaxies' of the Deutsche
Forschungsgemeinschaft (DFG). EB is grateful to CONACyT for financial
support through grant Nr.\ 27606-E. This research has made use of the
NASA/IPAC Extragalactic Database (NED) and the NASA/IPAC Infrared
Science Archive, which are maintained by the Jet Propulsion
Laboratory, Caltech, under contract with the National Aeronautics and
Space Administration (NASA), NASA's Astrophysical Data System Abstract
Service (ADS), NASA's SkyView, and the astronomical database SIMBAD,
provided by the 'Centre de Donn\'ees astronomiques de Strasbourg'
(CDS).

\bsp

\label{lastpage}

\end{document}